\documentclass[aps,prx,showpacs,showkeys,twocolumn,superscriptaddress,
floats,floatfix,preprintnumbers,longbibliography]{revtex4-1}
\usepackage{amssymb,amsmath}
\usepackage{bm,url,graphics,subfigure,epsfig,rotating,float}
\usepackage{mathrsfs} 
\usepackage{color,soul,graphicx,dcolumn,enumitem}
\usepackage{ulem}
\usepackage{lipsum}
\graphicspath{{./}}
\def \dd   {\rm{d}}
\def \rr   {\bm{r}}
\def \kaph {\hat{\kappa}}
\def \Ri {R_{\rm i}}
\def \Rj {R_{\rm j}}
\def \RRi {\bm{R}_{\rm i}}
\def \RRj {\bm{R}_{\rm j}}

\def \gdot {\dot{\gamma}}
\def \Kkappa {\bm {\kappa}}

\def \yy {\bm y}

\def \ri {\rm i}

\def \rr {\bm{r}}

\def \xhat {\hat{\rm  x}}

\def \XX {{\bm{X}}}

\def \Kbb {\mathbb{K}}
\def \ri {{\rm i}}
\def \rj {{\rm j}}
\def \rk {{\rm k}}

\def \del2z {\partial^{2}_{z}}

\def \dab {\delta^{\alpha\beta}}

\def \ab {\alpha\beta}

\def \uuh {\hat{\bm {u}}}

\def \UU {{\bm U}}

\def \delt {\partial_t}

\newcommand{\eq}[1]{(\ref{#1})}

\newcommand{\bra}[1]{\left\langle #1\right\rangle}

{}

\def \cP {\mathcal{P}}
\def \cQ {\mathcal{Q}}

\def \cF {\mathcal{F}}
\def \cJ {\mathcal{J}}
\def \cN {\mathcal{N}}

\def \RR {\bm{R}}
\newcommand{\fig}[1]{Fig.~\ref{#1}}
\newcommand{\subfig}[2]{Fig.~\ref{#1}(#2)}
\newcommand{\Fig}[1]{Figure.~\ref{#1}}

\def\drawing #1 #2 #3 {
\begin{center}
\setlength{\unitlength}{1mm}
\begin{picture}(#1,#2)(0,0)
\put(0,0){\framebox(#1,#2){#3}}
\end{picture}
\end{center} }

\def \mG {\mathcal{G}}

\def \mub {\overline{\mu}}

\def \mH {\mathcal{H}}

\def \eg {e\/.g\/. }
\def \ie {i\/.e\/. }

\def \etal {et.\ al.\ }
\def \Rij {\bm{R}_{\rm ij}}

\def \Ra {R^\alpha}
\def \Rb {R^\beta}

\def \hamil {\mathcal{H}}

\def \th {\rm {\th}}
\def \rp {{\rm p}}

\def \rk {{\rm k}}
\def \Np {N_{\rm p}}
\def \brho {\bm{\rho}}
\def \delttl {\partial_{\tilde{t}}}
\def \Rtl 	{\widetilde{R}}
\def \Ritl	{\Rtl_{\rm i}}
\def \Rjtl	{\Rtl_{\rm j}}
\def \Mtl	{\widetilde{M}}
\def \Rijtl	{\Rtl_{\rm ij}}
\def \mHtl	{\widetilde{\mH}}
\def \ttl 	{\tilde{t}}
\def \ytl 	{\tilde{y}}

\begin{document}
\title[]{Chaos and irreversibility of a flexible filament in periodically--driven Stokes flow}
\author{Vipin Agrawal}
\email{vipin.agrawal@su.se}
\affiliation{ Nordita, KTH Royal Institute of Technology and
Stockholm University, Roslagstullsbacken 23, 10691 Stockholm, Sweden.}
\affiliation{Department of Physics, Stockholm university, Stockholm,Sweden.}
\author{Dhrubaditya Mitra}
\email{dhruba.mitra@gmail.com}
\affiliation{Nordita, KTH Royal Institute of Technology and
Stockholm University, Roslagstullsbacken 23, 10691 Stockholm, Sweden}
\date{\today}
\begin{abstract}
The flow of Newtonian fluid at low Reynolds number is, in general, regular and time-reversible 
due to absence of nonlinear effects. 
For example, if the fluid is sheared by its boundary motion that is subsequently reversed, 
then all the fluid elements return to their initial positions. 
Consequently, mixing in microchannels happens solely due to molecular diffusion and 
is very slow.  
Here, we show, numerically,  that the introduction of a single, freely-floating, 
flexible filament in a time-periodic linear shear flow can break reversibility 
and give rise to chaos due to elastic nonlinearities, if the bending rigidity of the 
filament is within a carefully chosen range.  
Within this range, not only the shape of the filament is  spatiotemporally chaotic, 
but also the flow is an efficient mixer.
Overall, we find five dynamical phases: the shape of a stiff filament is 
time-invariant -- either straight or buckled;  it undergoes a period-two bifurcation 
as the filament is made softer; becomes spatiotemporally chaotic for even softer 
filaments but, surprisingly, the chaos is suppressed if bending rigidity is decreased further. 
\end{abstract}
\maketitle
\section{Introduction}
Flows at very small Reynolds number play a key role in
biology~\cite{purcell1977life,lauga2009hydrodynamics,phillips2012physical,
  taylor1951analysis} and
microfluidics~\cite{squires2005microfluidics, kirby2010micro,stone2004engineering}. 
In the presence of rigid boundaries, such flows are typically laminar and reversible.
For example, consider the fluid between two concentric cylinders sheared by rotating
the outer one slowly. 
If the rotation is reversed the fluid particles come back to their
original positions (ignoring the small fluctuations due to Brownian
motion)~\cite{taylor1967film}. 
Consequently, mixing by periodic stirring is in general
catastrophically slow in microfluidic flows although 
Lagrangian chaos is possible in pressure--driven flows through 
rigid but complex boundaries~\cite{aref2017frontiers}.
Addition of elastic polymers~\cite{groisman2000elastic,groisman2001efficient,
  groisman2000elastic,stroock2002chaotic,steinberg2021elastic},
or active objects~\cite{dombrowski2004self,wensink2012meso,dunkel2013fluid}
and mutual hydrodynamic interaction between many suspended colloidal
particles~\cite{pine2005chaos},
can also give rise to breakdown of time-reversibility and to chaotic 
flows.

Here we consider a neutrally--buoyant inextensible, elastic filament,
of length $L$ and bending rigidity $B$,
subject to a linear shear flow
$\UU_{0}(x,y) = \gdot y \xhat$.
The strain-rate $\gdot$ is time-periodic with a period $T$,
$\gdot = S\sin(\omega t)$, where $\omega = 2\pi/T$. 
Initially the filament is placed along the $y$ direction,
see \Fig{fig:snapshot}. 
The flow parameters, $S$, $T$, and dynamic viscosity of the fluid, $\eta$,
are chosen such that the Reynolds number is very small.
This corresponds to, for example, the demonstration by G.I. Taylor
where he puts a small thread in a Taylor-Couette apparatus filled
with syrup, turns the outer cylinder in one direction and then reverses it
exactly back to its starting position~\cite{taylor1967film}
This experiment does break time-reversal invariance -- the thread
is buckled at the end of the cycle. 
Here our aim is to study the same phenomena in a numerical setup
For simplicity, we consider a plane Couette flow without boundaries.

An elastic filament, of length $L$, in a constant-in-time flow 
has been extensively studied, numerically and 
experimentally~\cite{becker2001instability,
guglielmini2012buckling,liu2018morphological,lagrone2019complex,
slowicka2019flexible,zuk2021universal,kuei2015dynamics,hu2021levy,
chakrabarti2020flexible} for the last two decades.
Depending on the flow, the filament either attains
a complex shape, which is one case can even be 
helical~\cite{chakrabarti2020flexible}, or 
shows a wide range of morphological transition~\cite{liu2018morphological}
depending on its elastoviscous number $\mub \equiv (8\pi\eta SL^4)/B$
For small elastoviscous number (large bending rigidity), typically, 
the filament behaves like a rigid one. 
Hence we expect that in our case, if the bending rigidity 
of the filament is very large, the filament would rotate away and 
back to its original position without any change in shape.
We also expect that once the  bending rigidity is below a threshold,
(or $\mub$ is above a threshold)
the filament would buckle -- it would not return to its original shape.
The time reversibility would be broken.
If the bending rigidity is decreased further we expect
elastic nonlinearities to play a more and more dominant role
in the dynamics thereby giving rise to complex morphologies. 
Repeating the experiments over many cycles can potentially give rise to
spatiotemporaly chaotic behaviour of the filament. 
\section{Model}
We use the bead-spring model for the numerical simulation of the filament in
a Stokes flow \cite{larson1999brownian,wada2006non,guglielmini2012buckling,
  nazockdast2017fast,slowicka2019flexible,zuk2021universal}.
The model consists of $N$ spherical beads of diameter $d$,
connected by overdamped springs of equilibrium length $a$. 
The equation of motion for the $i$-th bead is given by~\cite{wada2006non}: 
\begin{subequations}
  \begin{align}
  \delt \Ri^{\alpha} &= -\sum_{j=0}^{N-1}M_{\rm ij}^{\alpha\beta}(\Rij)
  \frac{\partial \mH}{\partial \Rj^{\beta}}
  + U_0^\alpha(\RRi) \quad\/, \label{eq:dRdt}\\
  M_{\rm ij}^{\alpha\beta}(\RR) &=
  \frac{1}{8\pi\eta R}\Big[\dab + \frac{\Ra\Rb}{R^2}\nonumber \\
   &+\frac{d^2}{2R^2}\left(\frac{\dab}{3} -
    \frac{\Ra\Rb}{R^2}\right)\Big], \quad \text{for}\quad  \ri \neq \rj
            \nonumber \\
            &=\frac{1}{3\pi\eta d} \dab \quad\text{for}\quad \ri = \rj
            \label{eq:RP}
\end{align}
\end{subequations}
Where $\RRi$ is the position vector of the center of the $i$-th bead,  
$\Rij \equiv \RRj - \RRi$,
$\UU$ is the velocity of the background shear flow,
and $\eta$ is the dynamic viscosity of the fluid.

The hydrodynamic interaction between the beads is described by
the Rotne--Prager mobility tensor $M_{\rm ij}(\RR)$~\cite{rotne1969variational,
  brady1988stokesian, guazzelli2011physical,kim2013microhydrodynamics}
derived by solving the Stokes equation,
i.e., our simulations are at zero Reynolds number. 
The Latin indices run from $1$ to $N$, the total number of beads, and the
greek indices run from $1$ to $D$, where $D=3$ is dimensionality of space.
Repeated greek indices are summed.

The elastic Hamiltonian, $\mH$,
contains contribution from both bending and
stretching but not torsion:
  $\mH = \mH_{\rm B} + \mH_{\rm S}$,
where $\mH_{\rm B}$ and $\mH_{\rm S}$ are contributions from
bending \cite{montesi2005brownian,bergou2010discrete} and
stretching \cite{wada2006non,wada2007stretching} respectively.
The bending energy of a continuous filament is given by \cite{powers2010dynamics}:
\begin{equation}
  \mH_{\rm B} = \frac{B}{2}\int \kappa^2(s) \dd s,
\end{equation}
where $B$ is the bending modulus, $\kappa$ is curvature, and $s$ is
the material coordinate.
As we use a discrete bead--spring model, hence we must discretize
the Hamiltonian, see appendix~\ref{sect:method}.
%For simplicity, we consider a practically inextensible filament -- implemented by
%choosing an appropriately high stretching modulus --  with
%only bending but no torsion
%~\footnote{The length of the filament changes
%  by at most $2\%$ in the worst case.}.

We define three dimensionless parameters:
the elastoviscous number, $\mub$, the non-dimensional frequency, $\sigma$,
and the ratio of stretching to bending, $K$, defined respectively
as
\begin{equation}
  \mub = \frac{8\pi\eta S L^4}{B},\quad
  \sigma = \frac{\omega}{S},\quad\text{and}\quad
  K = \frac{H d^2}{B}\/.
\end{equation}
The elastoviscous number measures the relative strength of the elastic
forces due to bending and the visous forces.

In appendix~\ref{sect:method}, 
we give a 
comprehensive description of the model,
its numerical implementation, and 
the parameters of simulations.
The elastoviscous number of our simulations includes  in the same
range as the experiments in Ref.~\cite{liu2018morphological}. 
Our code reproduces their experimental results, see appendix~\ref{sect:method}.

The velocity of the flow at any Eulerian point $\rr = (x,y,z)$ is given by
\begin{subequations}
  \begin{align}
    U^{\alpha}(\rr) &= U_0^{\alpha}(\rr) + \mG^{\ab}(\rr-\RRi)F^{\beta}_{\rm i}\/,\quad\/,
      \text{where}\label{eq:Ueuler}\\
      F^{\alpha}_{\rm i} &= -\frac{\partial \hamil}{\partial R^{\alpha}_{\rm i}}\/,
      \quad\text{and}\label{eq:ff}\\
      \mG^{\ab}(\RR) &= \frac{1}{8\pi\eta R}\left[ \dab + \frac{\Ra\Rb}{R^2}
        +\frac{d^2}{4R^2}\left( \frac{1}{3}\dab - \frac{\Ra\Rb}{R^2} \right) \right]
      \label{eq:Green}
  \end{align}
  \label{eq:tracervel}
\end{subequations}
In \eq{eq:Ueuler} $U_0^{\alpha}(\rr)$ is the background linear
  shear flow and $\mG^{\ab}$ is the Green's function given in
\eq{eq:Green}. 
%--------------------------------------------%
\section{Results}
%----------------------------------------------------------------
\begin{figure}
  \includegraphics[width=0.9\columnwidth]{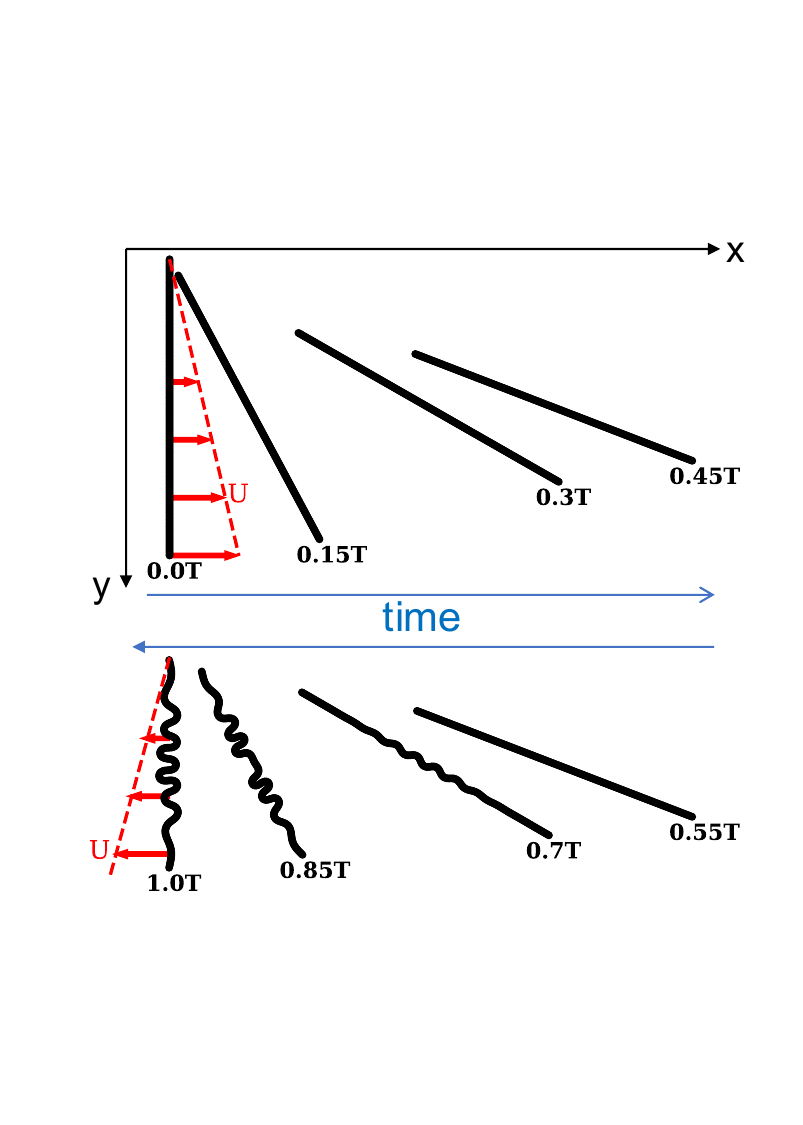}
  \caption{{\bf Sketch of numerical experiment}: Initially the filament is
    straight. It rotates and translates while advected by 
    $\UU = \gdot y \xhat$ with $\gdot = S\sin(\omega t)$
    over the first-half of the cycle.
    In the second half the filament rotates and translates back
    but in addition buckles if its elastoviscous number is large enough.
    The flow $\UU$ at $t=T/4$ (top panel) and $t=3T/4$ (bottom panel)
    are shown as red arrows.
    \label{fig:snapshot}}
\end{figure}
%--------------------------
\begin{figure*}
  \begin{center}
    \includegraphics[width=\textwidth]{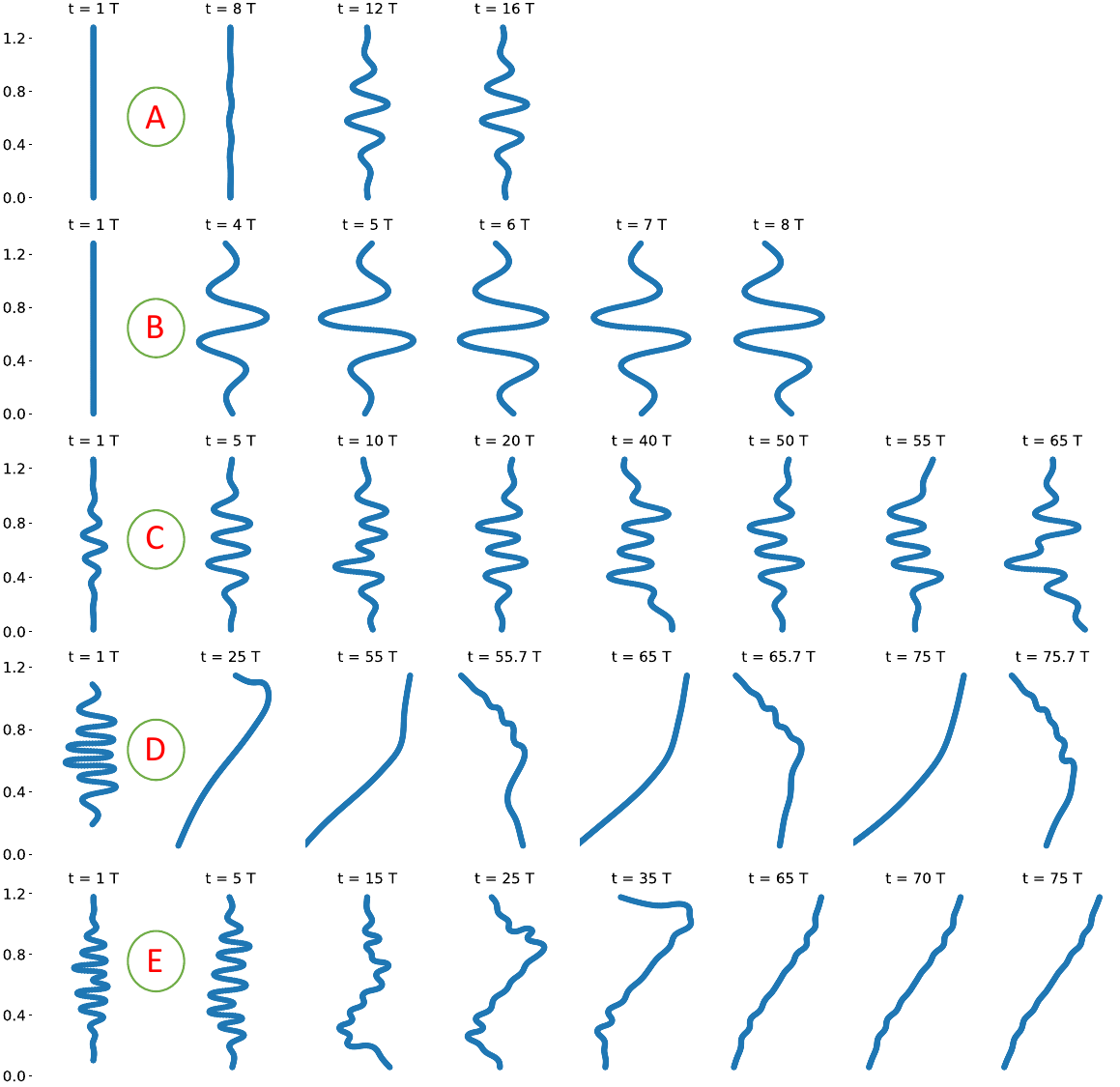}
    \caption{{\bf Kaleidoscope of dynamical behavior}:
      (A) {\bf Periodic buckling} ($\mub = 1.46 \times 10^6$, $\sigma = 1.5$):
      The filament is buckled after $8$ cycles and repeats
      itself after every cycle afterwards. Till cycle $8$ the filament is not straight but
      slightly deformed from its initial straight shape. This small deformation is
      barely noticeable in this figure.  
      (B) {\bf Two-period} ($\mub = 0.67 \times 10^6$, $\sigma= 0.75$):
       The shape of the filament is either of two shapes, which are mirror images of 
       each other, at odd and even cycles.
      (C) and (D) The two types of {\bf Complex} phases:    
      (C)For $\mub = 3.35 \times 10^6$, $\sigma = 1.5$: 
        The filament never repeats itself.  
      (D) For $\mub = 6.7 \times 10^6$, $\sigma = 0.75$: The filament shows
      spatiotemporally complex behavior but for late cycles, 
      the filament (almost) repeats at the end of every cycle
      ($nT$) ($t=55T,65T,75T$) but not at any other time. To illustrate, we 
      show the snapshots at  $t=55.7T,65.7T,75.7T$ -- the shape of the filament
      is different from one another.  
      Note, the filament shows maximum buckling, not at the end of a cycle, but 
      at times in-between the cycles, i.e., at  $t=(n+p)T$, where $n$ is an integer and $0<p<1$.
      (E) {\bf Complex transients}($\mub = 16.75 \times 10^6$, $\sigma = 1.5$):
      Filament shows complex behavior for early periods but repeats itself at late times. 
    \label{fig:trans}}
  \end{center}
\end{figure*}
%---------------------------------------
%-------------------------------
\begin{figure}
  \includegraphics[width=1\columnwidth]{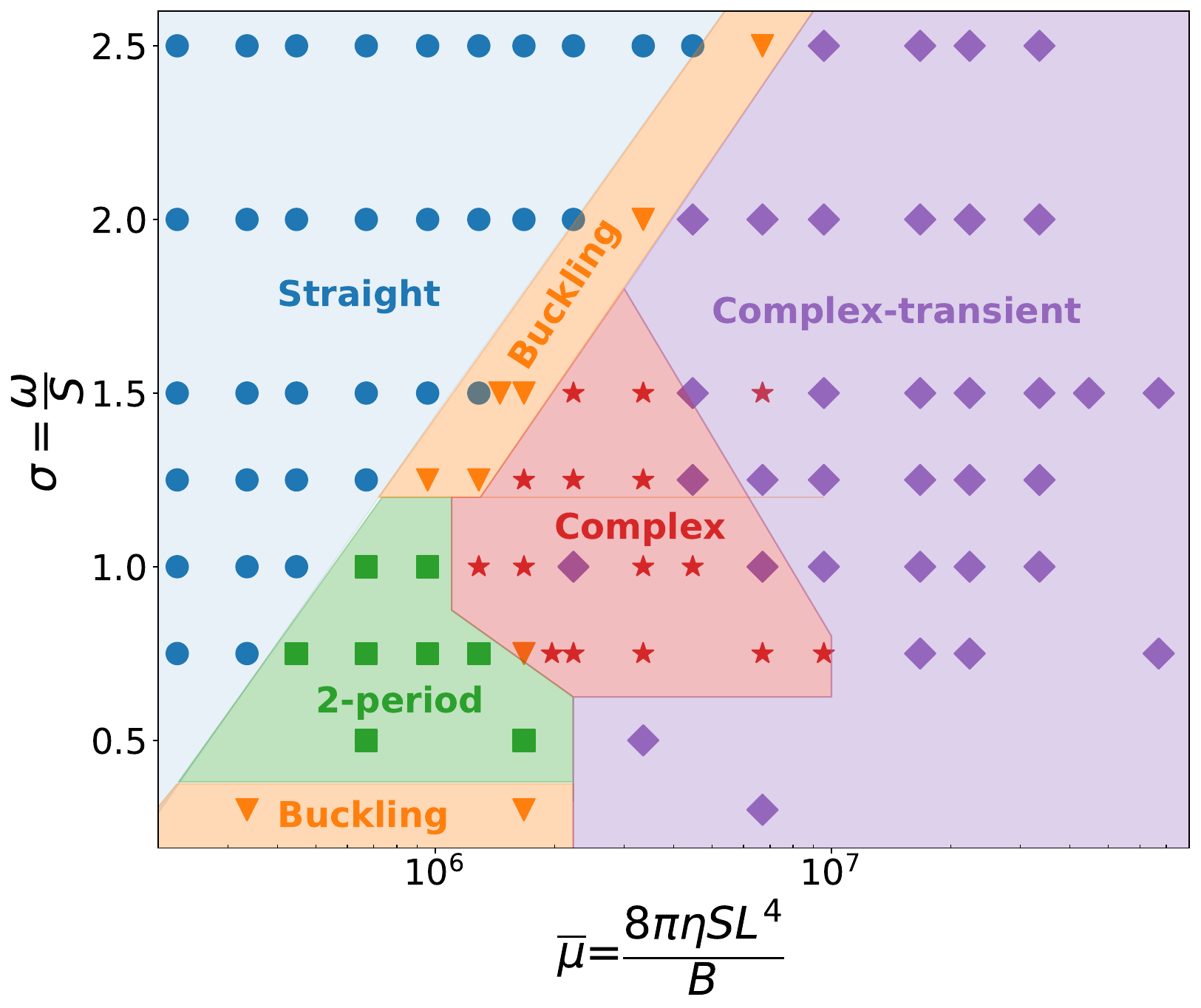}
  \caption{
    {\bf Phase diagram} from time-dependent numerical simulations in the
      $\mub$--$\sigma$ parameter space;
      $\left(\sigma = \frac{\omega}{S},
      \mub = \frac{8\pi\eta S L^4}{B}\right)$.
      Here $\omega$ is rate of change of strain, $S$ is strain rate,
      $\eta$ is the viscosity, $L$ is length of the filament,
      and $B$ is the bending modulus.
      Initially, the filament is freely suspended in the shear flow, see \Fig{fig:snapshot}.  
      We show five different dynamical phases in the system represented by five symbols.
       {\bf Straight} ($\bullet$) : The filament comes back to the initial position in
      the straight configuration after every period.
      {\bf Periodic buckling} ($\blacktriangledown$):- The filament comes back in the buckled
      configuration after every period.
      {\bf Two-period ($\blacksquare$)} : The filament repeats its configuration not after
      every but after two-period.
      {\bf Complex}($\bigstar$): The filament buckles into complex shape with very high
      mode of buckling instability. 
      {\bf Complex--transients}($\blacklozenge$) : Filament shows long transients with complex
      shape but at late times, the shape of the filament repeats itself.
    The boundary between the complex and complex-transient phase is difficult to
    clearly demarcate.  
    \label{fig:phase}
    }
\end{figure}
%---------------------------------------
Initially, the  filament is placed along the $y$ direction
with its head at $y=0$, see \Fig{fig:snapshot}. 
We use $N=256$, $K=16$ and $a=d$ in all our simulations
and vary both $\sigma$ and $\mub$ to explore a variety of dynamical
behaviour.

As we impose an external linear flow with a period $T$,
it is useful to look at \textit{stroboscopic} snapshots of the filament separated 
by time $T$.
We note that in some cases filament does not return to its original position
at the end of a cycle.
When that happens time-reversal invariance is already broken even if the shape of
the filament remains unchanged. 
We call this \textit{swimming}.
In this paper we focus not on swimming but on how the shape of the filament changes. 
\subsection{Dynamical phases}
Based on the shape, a kaleidoscope of qualitative behavior emerges that
we classify into five  different dynamical phases.
For small elastoviscous number ($\mub$) the filament comes back to its original
position undeformed at the end of every period (not shown in \fig{fig:trans}).
As $\mub$ is increased, the filament is buckled at the end of every
period; see \fig{fig:trans} panel A.
On increasing $\mub$ further we reach a phase where two buckled states,
which are mirror images of each other, alternate at the end of even and odd periods
-- \textit{a period--two solution}; see \fig{fig:trans} panel B.
At even higher $\mub$ the shape of the filament at the end of every cycle
is different -- the filament never repeats itself.
We continue these simulations to long times.
In some cases the shape of the filament never repeats itslf,
see \Fig{fig:trans} panel C.
In some other cases, the shape of the filament is almost repeated
at the end of every period, but the shape is different at all other times,
e.g., consider the snapshots in \Fig{fig:trans} panel D.
The shape of the filament at $65T$ and $75T$ are almost the same.
But the shapes at $65.7T$ and $75.7T$ are not. 
We do not make a distinction between these two dynamical behavior
-- we call both of them complex. 
Finally, for large $\mub$ we find that the shape of the filament shows
complex spatiotemporal behaviour till intermediate times $t < 60 T$ but
almost settles (it comes very close but does not repeat itself) to a fixed spatially complex shape at late
times; see \fig{fig:trans} panel E.
By surveying a range of values for $\mub$ and $\sigma$ we construct a phase
diagram in \fig{fig:phase}.
We find that the straight phase can go unstable in two ways,
depending on the value of $\sigma$.
It can either undergo a bifurcation to a two-period solution or
go to a buckled phase which repeats itself.
The buckled solution appears at end of every period, it is a time-reversible solution,
the two-period solution is not.
The boundary between the phases can be clearly demarcated except the boundary
between the complex and the complex-transient phase.
Thus it may be possible that there is  a fractal boundary between these
two phases.
Fractal boundaries are not an uncommon occurance in many dynamical systems
including transition to turbulence in pipe flows~\cite{schneider2007turbulence},
different forms of spiral-wave dynamics in mathematical models of cardiac
tissues~\cite{shajahan2007spiral},
and onset of dynamo in shell-models~\cite{sahoo2010dynamo}.
We have not explored this aspect in any detail in this paper. 
Except the region of the phase diagram where we find straight solutions,
swimming solutions appear everywhere else. 
\subsubsection{Complex Phase}
Let us first discuss in detail a representative simulation in the complex phase.
As we focus on the shape of the filament we describe the filament in its
intrinsic coordinates -- its curvature ($\kappa$) as a function of
material coordinate ($s$).
We calculate curvature using a discrete approximation, see
  appendix~\ref{sect:method}.
In  \fig{fig:CP_curv} we plot $\kappa$ versus $s$
for early times, at $t = T, 10T, 19T$ and $28T$,  in \subfig{fig:CP_curv}{A}
and for
late times in \subfig{fig:CP_curv}{C}, at t=$35T, 45T, 55T$ and $75T$.
At all times, the curvature is zero at the two end of the filament, as
dictated by the boundary conditions, and changes sign several times,
i.e., a quite complex morphology is observed, as we show in
\subfig{fig:CP_curv}{B} and \subfig{fig:CP_curv}{D} respectively.
The minimum value of the radius of curvature is approximately $10 d$
where $d$ is the diameter of each bead or alternatively the thickness of the
filament. 
A sine transform of the $\kappa(s)$ to $\kaph(q)$  shows several peaks,
For $q \gtrapprox 20$ the $\kaph$ is practically zero,
see Appendix~\ref{sect:phase}.
This demonstrates that our simulations are well resolved to capture the
phenomena we observe. 
  As the filament moves in the fluid it changes the background flow.
  In \subfig{fig:CP_curv}{E} and \subfig{fig:CP_curv}{F} we plot a typical
  phase-potrait of the velocity
  of the flow (after subtracting out the background velocity) at a fixed Eulerian
  point.
  For small $\mub$, in the part of the phase diagram where the
  filament always remain straight, the phase potrait is a simple closed curve as shown
  in orange in \subfig{fig:CP_curv}{E} and \subfig{fig:CP_curv}{F}.
  For the case where the filament is in the complex phase we obtain
  a non-trivial attractor.

There is another intriguing feature seen in some of the runs in the complex phase:
although the filament never repeat itself exactly in the $\kappa$--$s$ space
it comes very close to periodic behavior with a large period, in one
case $30T$. 
\subsubsection{Complex--Transient phase}
Next we turn to the phase we call \textit{Complex--Transient}.
Here the behavior is the same as the \textit{Complex} phase upto quite late
times, e.g., $60T$ after which the filament comes to almost the same
shape at the end of every period.
Here also the dynamics of the filament is not strictly periodic.
The shapes, which are complex,  change but very slowly over time. 
This slow drift in the configuration space can be measured by calculating
\begin{equation}
  K(p-m) = \left[ \int\mid\kappa(s,pT)-\kappa(s,mT)\mid^2 ds \right]^{1/2}
  \label{eq:K}
\end{equation}
where $m > 60$ is a period where the filament has already reached its late time
behavior. 
We find $K(p-m) \sim (p-m) $, for not too large values of $p-m$.  i.e.,
an algebraic growth. 
We perform another numerical experiment. We take the filament
in its late almost stationary configuration and add a small perturbation
and then evolve again.
The perturbation goes to zero very quickly, the filament goes back to its
almost stationary configuration. 
%-------------
\begin{figure*}
  \includegraphics[width=0.95\linewidth]{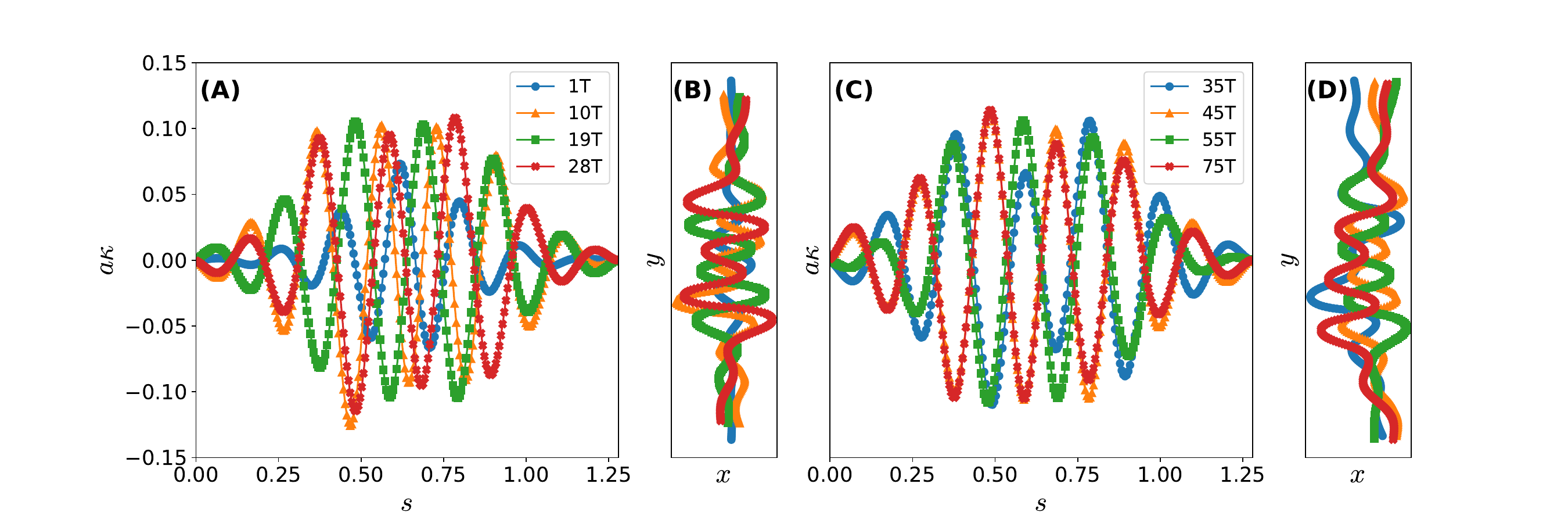}
  \includegraphics[width=0.95\linewidth]{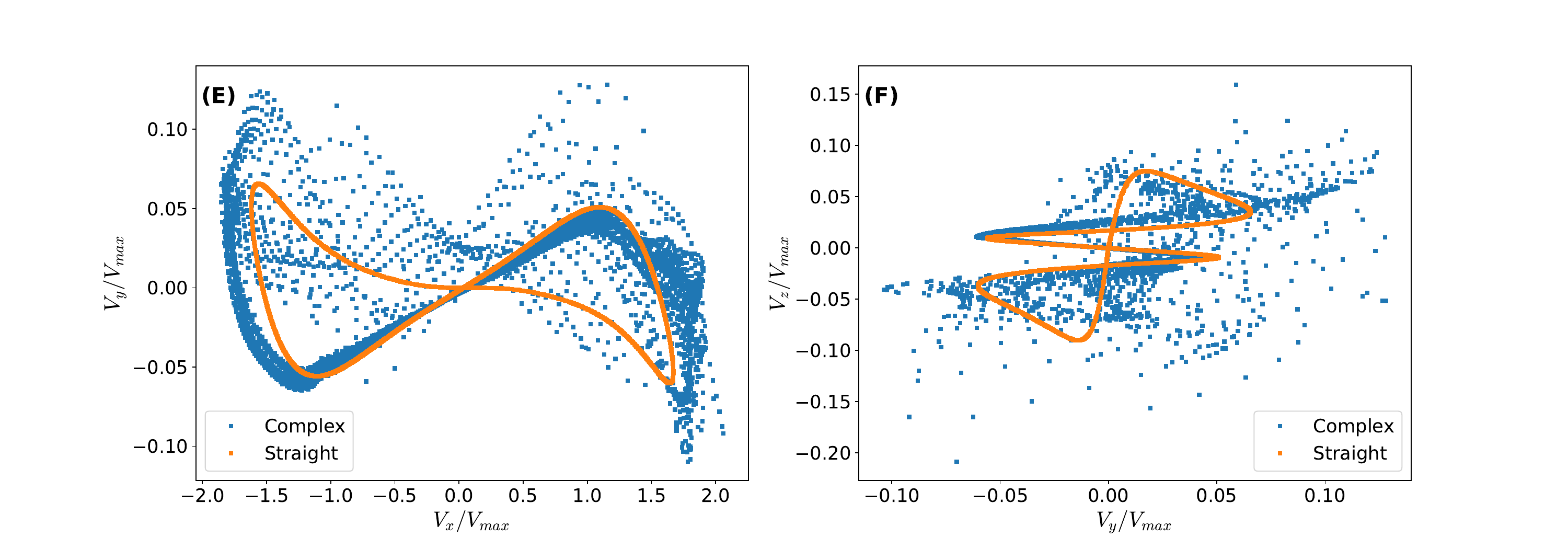}
    \caption{{\bf Evolution in the complex phase}: 
      ($\mub = 3.35 \times 10^6$, $\sigma = 1.5$) 
      The filament shows complex behavior and does not repeat in real space or
      configuration space ($t < 45 T$) \ref{fig:ComplexPhaseEnd}.
      (A) Curvature ($\kappa$) as a function of material coordinate, $s$, of 
      the filament at early cycles, $t=T, 10T, 19T$ and $28T$.  
      (B) Image of the filament at the same times. 
      (C) and (D) Same as (A) and (B) but for late times.  
      (E,F) Phase portrait of tracer velocity at a fixed Eulerian point for late cycles
      $t = 45T$--$75T$.}
    \label{fig:CP_curv}
\end{figure*}
%------------------------------
\subsection{Stroboscopic map}
So far we have studied the different dynamical phases through
time-stepping our numerical code.
Potentially, both the complex and complex-transient phase are spatiotemporally
chaotic.
But a time-stepping code however accurate accumulates error which grows with
the number of time-steps taken.
To investigate the fate of the filament at late times we do have to integrate
over long times.
Hence we need additional evidence to confirm the existence of chaos in this
problem.

We start by defining the stroboscopic map, $\cF$,
\begin{equation}
\kappa(s,T) = \cF \kappa(s,0) 
\end{equation}
The dynamical system obeyed by the filament, equation \eq{eq:dRdt}, is non-autonomous
because the external flow $\UU$ is an explicit function of time, the
map $\cF$, which is generated by integrating \eq{eq:dRdt} over exactly one
time period $T$, is autonomous.
This is a map of $N$-dimensional space onto itself where $N$ is the number of beads.
The function $\kappa(s,t)$ at $t= nT$ and $t=(n+1)T$ are related by 
one iteration of this map.  
We proceed to study the fixed points and periodic orbits of this map
as a function of the parameters, $\mub$ and $\sigma$,
following Refs.~\cite{auerbach1987exploring,cvitanovic2005chaos}.
Such techniques have been used widely to study transition to turbulence
in high-dimensional
flows~\cite{kerswell2005recent,suri2020capturing,page2020exact}
and has also been applied to other fields of fluid
dynamics~\cite{franco2018bubble,gaillard2021life}.
The detailed numerical techniques are described in appendix~\ref{sect:method}.

In \fig{fig:GMRESPD} we show several examples of the solutions we obtain,
for a fixed $\sigma=0.75$ as a function of $\mub$.
For small $\mub = 0.17\times 10^{6}$ we obtain only one fixed point and it
corresponds to $\kappa(s)=0$, i.e., a straight filament.
At $\mub=0.33\times 10^{6}$ in addition to the straight filament a new fixed
point appears, where $\kappa$ is zero at one end, changes sign
once roughly at the middle of the filament and has two maximas.
We show the shape of the filament in \subfig{fig:GMRESPD}{B1}.
At exactly this point in the phase diagram, see top line in \fig{fig:GMRESPD},
obtained from the evolution code, we find a straight filament.
This implies, either of the two possibilities: one, the new solution is unstable;
two both the solutions are stable but the evolution code lands up in the
straight one because of the initial condition we chose. 
Next at $\mub = 0.33\times 10^{6}$ we no longer find any fixed points.
We find two periodic orbits, one that is a  two--period~\subfig{fig:GMRESPD}{C1}
and one with four periods~\subfig{fig:GMRESPD}{C2}.
The two solutions in the two-period solution are mirror images of each other.
At the same place in the phase diagram the evolution code
finds the same two--period solution.
Increasing $\mub$ to $0.84\times 10^{6}$ we find that the four-period solution
has disappeared, two two--period solutions exist, \subfig{fig:GMRESPD}{D}.
At even higher values of $\mub$ we start to obtain many solutions.
We show a few examples in \subfig{fig:GMRESPD}{F}, \subfig{fig:GMRESPD}{G},
and \subfig{fig:GMRESPD}{H}.
This is the region of phase space where complex and complex-transient dynamical
phases are seen. 

To summarize, by turning our system of coupled
non-autonomous differential equations, \eq{eq:dRdt} to an
autonomous stroboscopic map and by studying the solutions of the
map we find further
support of breakdown of time reversibiity and appearence of chaos
that we had already seen from the evolution of the differential equations.
We demonstrate that 
the first appearence of breakdown of time-reversibility is through
a period-two bifurcation.
The map has many solutions and the number of solutions increases as
we increase $\mub$.
%-------------------------
\begin{figure*}
    \includegraphics[width=\textwidth]{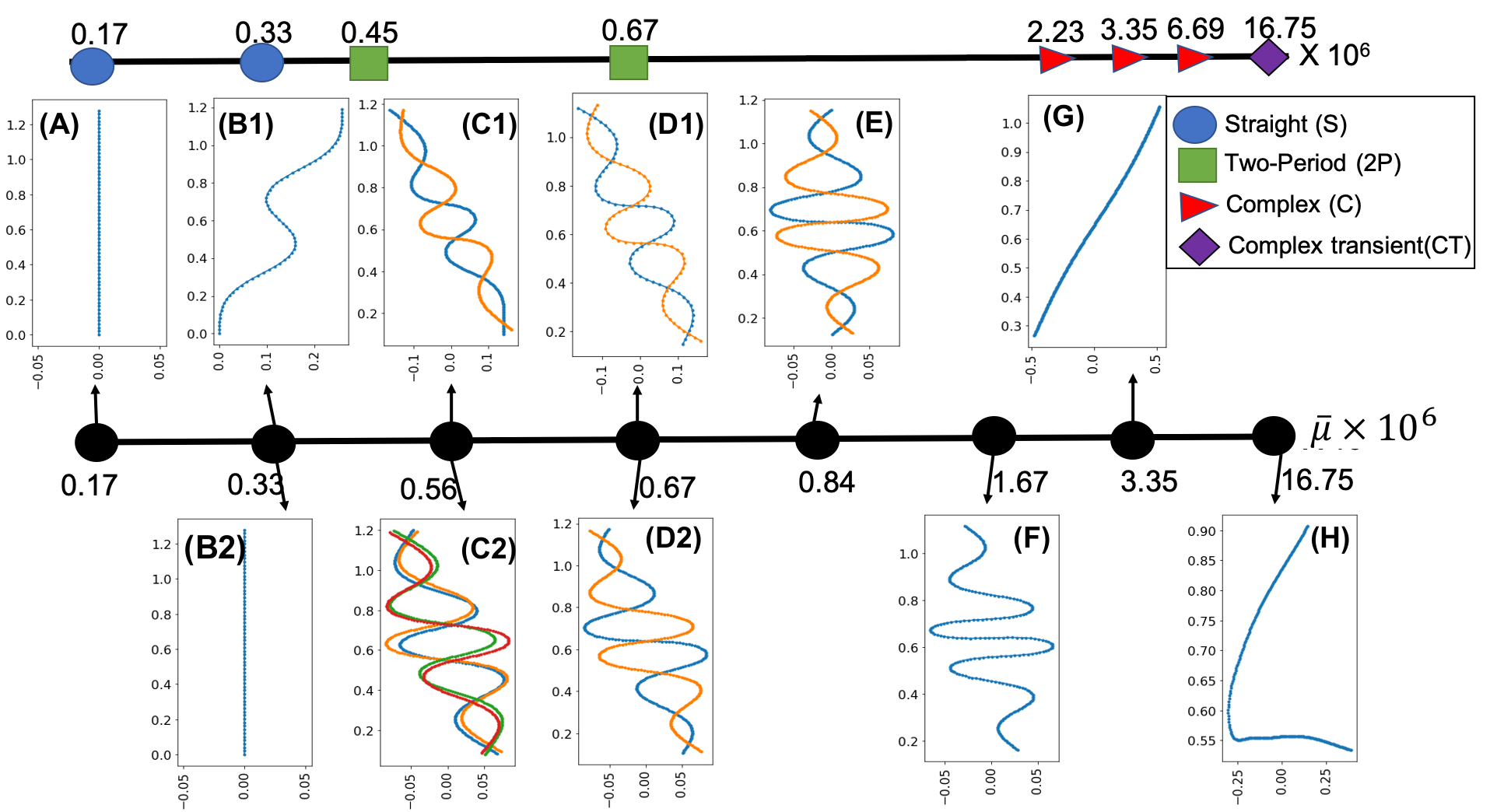}
    \caption{{\bf Solutions of stroboscopic map} in real space for $\sigma=0.75$ 
     for different values of $\mub$.  
      We find multiple co-existing solutions as we increase
      $\mub$ (black symbols from left to right) indicating the complexity of the system.
      We compare this solutions with the solutions obtained at late times from the
      evolution code  at the same points in the phase diagram.
    }
    \label{fig:GMRESPD}
\end{figure*}
%-----------------------------
\subsection{Mixing of passive tracer}
\begin{figure*}
  \includegraphics[width=0.32\linewidth]{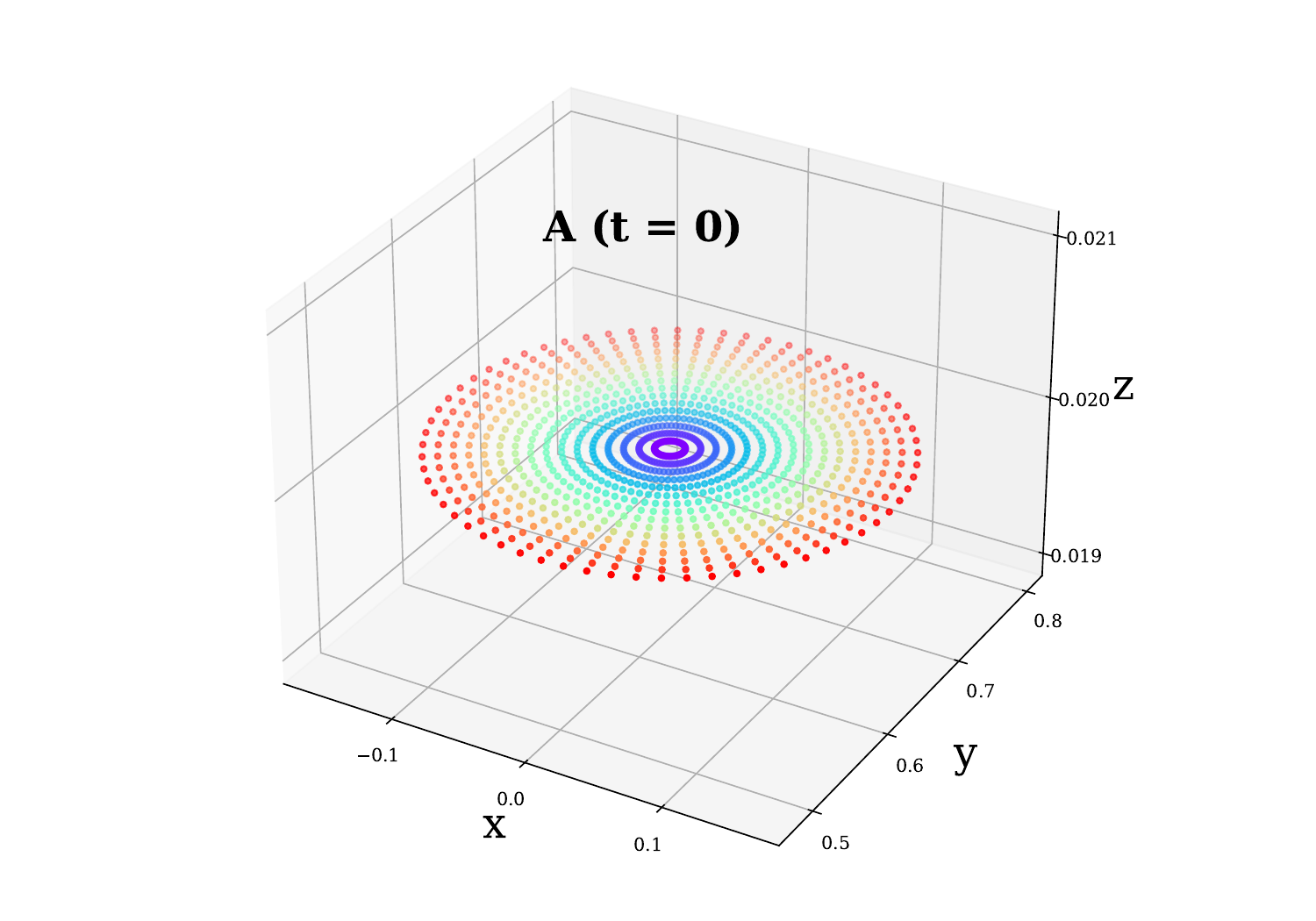}
  \includegraphics[width=0.32\linewidth]{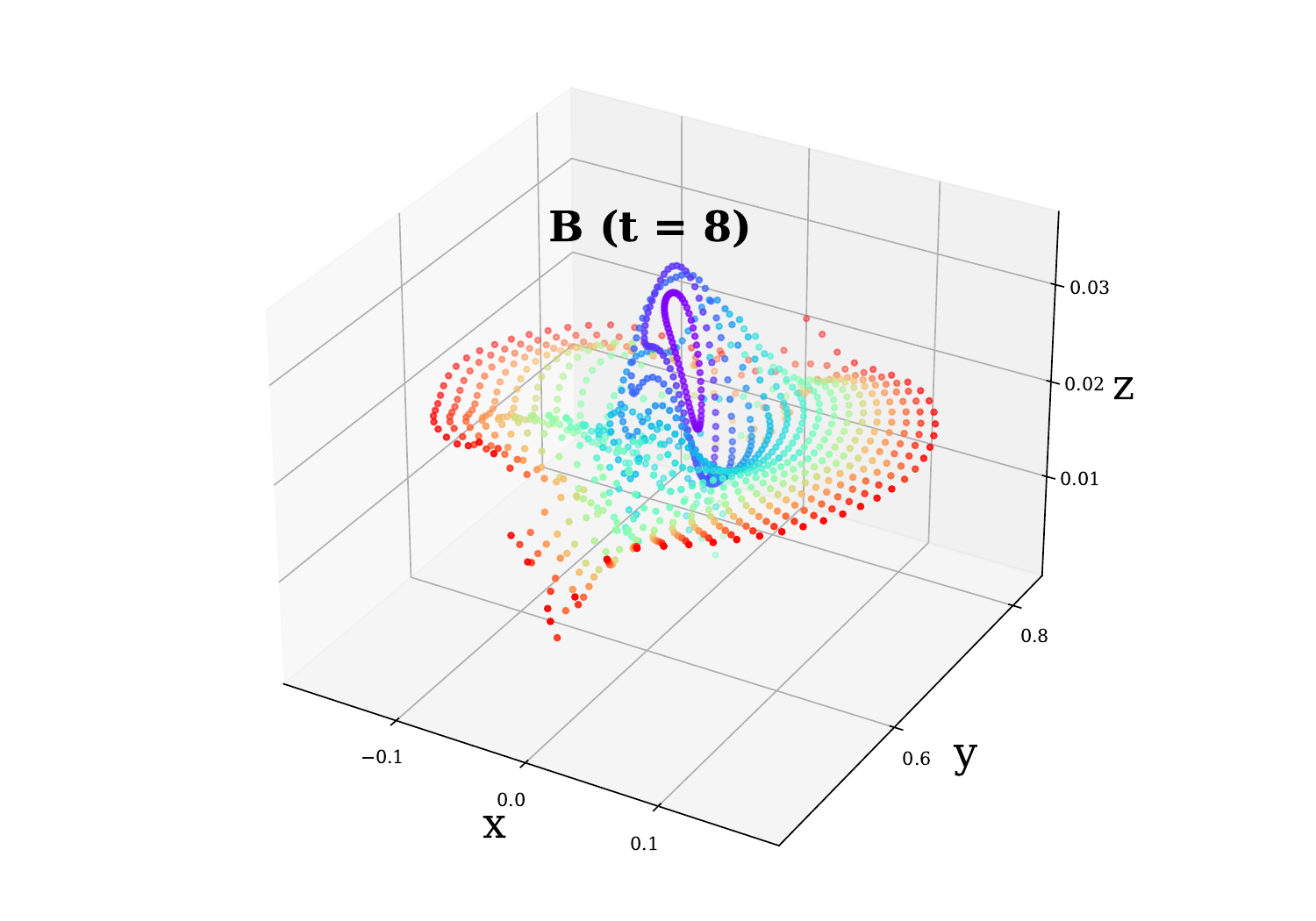}
  \includegraphics[width=0.32\linewidth]{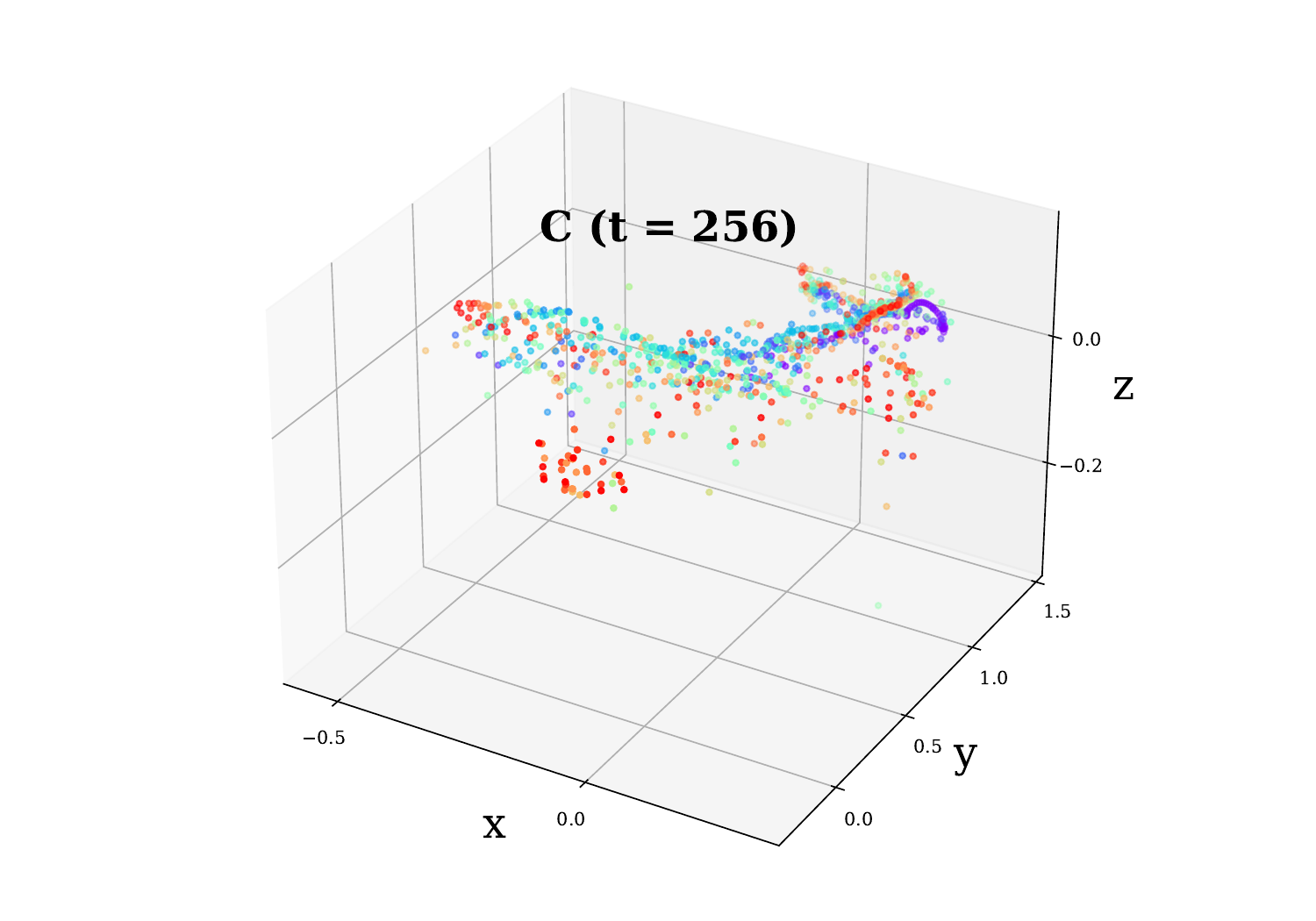}\\
  \includegraphics[width=0.45\linewidth]{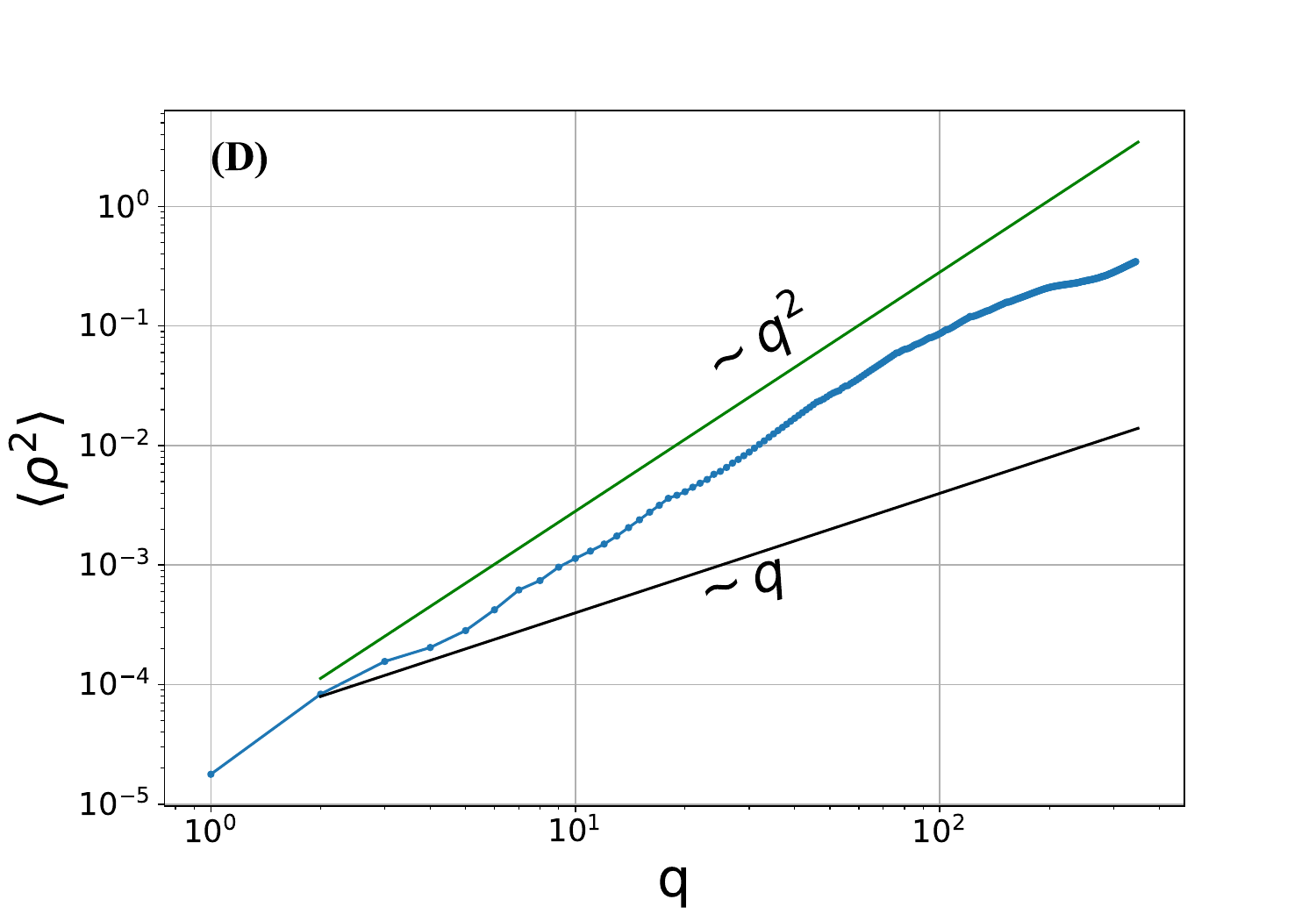}
  \includegraphics[width=0.45\linewidth]{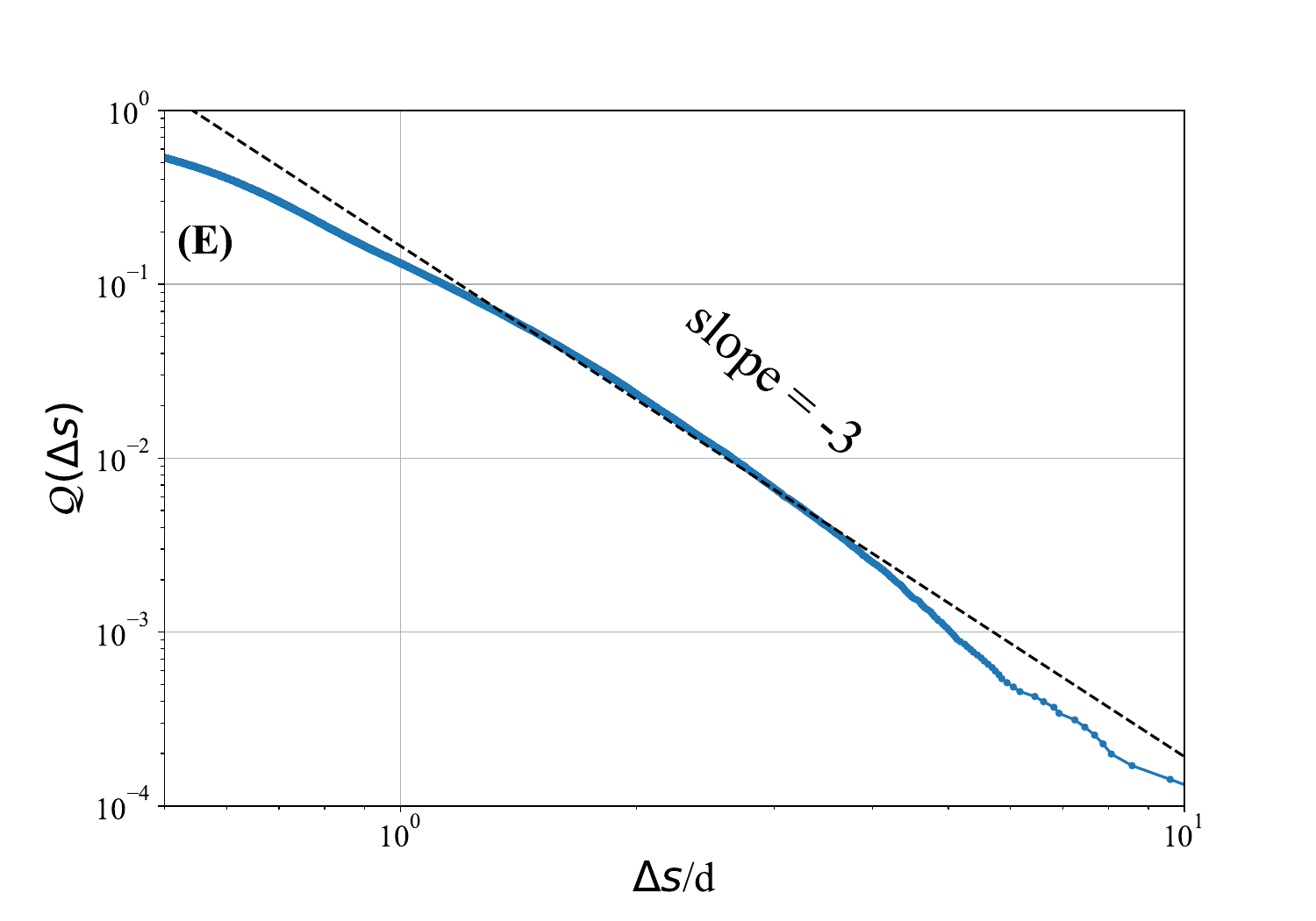}
  \caption{{\bf Mixing of passive tracer}:
    (A) to (C)  Positions of tracer particles  
    at different times for the filament in complex
    phase~($\mub = 3.3\times 10^6$, $\sigma=1.5$).
    Initially, the tracers are placed on concentric circles, color coded by their
    distance from the center of the circles.
    The mixing of the colors show the mixing of the scalars. 
   (D) Mean squre displacment (MSD), $\bra{\rho^2}(qT)$, defined in \eq{eq:msd},  
    as a function of $q$ in log-log scale.
    We also plot two lines with slopes $1$ and $2$. 
    (E) Cumulative probability distribution function, $1-\cQ(\Delta s)$ 
    as a function of
    $\Delta s = \sqrt{X_1^2 + X_2^2}$
    ($X_1$ and $X_2$ are in-plane coordinate of the tracers).
    }
    \label{fig:mixing}
\end{figure*}

Next we study how passive tracers are transported by the flow generated
by the presence of the filament
for parameters in the complex phase.
The velocity of the flow, $\UU(\rr)$, at any Eulerian point, $\rr = (x,y,z)$
is given in \eq{eq:Ueuler}.
The equation of motion of a passive tracer, whose position at time $t$ is given
by $\XX(t)$, is
\begin{equation}
  \frac{d\XX}{dt} = \UU(\rr)\delta(\rr-\XX)\/.
  \label{eq:tracer}
\end{equation}
We solve \eq{eq:tracer} simultaniously with \eq{eq:Ueuler} and for $\Np$ tracers.
The tracers are introduced into the flow on concentric circles
in the $x$--$y$ plane, \subfig{fig:mixing}{A}, after
approximately $10$ cycles, when the flow has reached a statistically stationary
They are colored by radius of the circle on which they lie on at the initial
time.
After $8$ periods, $t=8T$, we find that the outer rings are still somewhat
intact but the inner rings have somewhat merged with each other and also moved
out of the $x$--$y$ plane.
At even later time, $t=256T$, we find the tracer particles are well mixed with each
other.

In the rest of this section, we set $t=0$ at the time the tracers
are introduced. 
To obtain a quantitative measure of mixing we define
\begin{equation}
  \Delta \XX^{\rm k}_{\rm j} \equiv \XX^{\rm k}((j+1)T) - \XX^{\rm k}(jT)\quad\/,
  \label{eq:Dx}
\end{equation}
the net displacement of the $\rk$-th tracer particle over the
$j$-th cycle -- $t=jT$ to $t=(j+1)T$, where $j$ is
an integer.
The net displacement of the $\rk$-th tracer after $q$ cycles is
\begin{equation}
  \brho^{\rk}(q) = \sum_{j=1}^{q}\Delta \XX^{\rk}_{\rj}\/.
  \label{eq:rho}
\end{equation}
The total mean square displacement, averaged over all the tracers, at the end of
$q$ cycles is given by
\begin{equation}
  \bra{\rho^2(q)} \equiv \frac{1}{\Np}\sum_{\rk=1}^{\Np}
  \lvert \brho^{\rk}(q) \rvert^2 \label{eq:msd} 
\end{equation}
If the tracers diffuse then we expect $\bra{\rho^2(q)} \sim q$ for
large $q$~\cite{tay22}.
In \subfig{fig:mixing}{D} we plot $\bra{\rho^2(q)}$ versus $q$ in log-log
scale.
Clearly $\bra{\rho^2(q)}$ increases faster than $q$ but slower than $q^2$!
Could it be possible that the tracers show Levy-like superdiffusion ?

If this is true then the probability distribution function (PDF), $\cP$, of the
displacement $\Delta X^{\rk}_{\rj}$ must have a power-law tail with an
exponent $\gamma \geq -2$.
We probe this by calculating the cumulative probability distribution (CDF) function
for $\Delta X^{\rk}_{\rj}$ for all $j$ and $k$.
We calculate the CDF using rank-order method. 
The advantace of using the CDF is that unlike the PDF it is not plagued by binning
error. 
The CDF of $\Delta X^{\rk}_{\rj}$ is different for each of its components.
The CDF of the out-of-plane component, $X_3$, has an exponential tail. 
The CDF of the two in-plane components are qualitatively similar,
hence we calculate
$\Delta s = \sqrt{X_1^2 + X_2^2}$
and plot its CDF, $\cQ(\Delta s)$, calculated by the
rank-order method, in \subfig{fig:mixing}{E}.
The tail of the CDF has a slope approximately equal to $-3$, which implies that
the tail of the corresponding PDF has a slope of approximately $-4$.
Thus, by the central limit theorem we conclude that the tracers to show diffusion,
not superdiffusion.
However, as the PDF has power-law tail we expect that very long averaging
over very many number of tracer particles is necessary for convergence.
This explains why we do not observe clear evidence of diffusion from the
mean square displacement. 
%-------------------------------
%--------------------
\section{Conclusion}
To summarize, we numerically study the motion of a freely-floating elastic filament
in a linear shear flows  that changes periodically with time, at zero Reynolds number.
We find that elastic nonlinearies of the filament are responsible for
breakdown of time-reversal symmetry.
The first signature of this breakdown, which appears as we increase the
elasto-viscous number,
is that the filament starts to \textit{swim} --
it does not return to its initial poistion after one period, although it
returns to the same shape.
As the elastoviscous number is increased we find period-two bifurcation
and eventually what could be spatiotemporally chaotic behavior of the shape of the filament.
Surprisingly, at quite large elastoviscous number we find that chaos is suppressed at late times --
the filament returns to the same shape at the end of every period but
does not repeat itself between the periods.
We also demonstrate that such a filament is an efficient mixer of a passive scalar. 
Few comments are now in order.

Our numerical experiments corresponds, roughly, to the same range of elastoviscous
parameters as the recent experiments~\cite{liu2018morphological} of flexible filaments
in constant-in-time shear flow and our code reproduces the behaviour seen in these
experiments.
Hence we expect it will be possible to experimentally confirm our results,
at least qualitatively.
Intriguingly, the spatiotemporally chaotic behavior is observed within a window of
values of the elastoviscous number for a fixed value of $\sigma$. 

We have not confined the numerical solution of our problem to two dimensions.
The filament could, in principle, bend out-of-plane when buckled, but it never does.
We expect, if torsion is included the filament will bend out of the
plane and also break the reflection symmetry.
However, the passive tracers driven by the filament do move
out of the plane. 

In addition to elastic nonlinearity, we have included non-local
viscous interaction.
In its simplest approximation a filament in a viscous flow can be described
by including only the diagonal term ($\ri=\rj$) in the Rotne--Prager
tensor~\cite{goldstein1998viscous} in \eq{eq:RP}.
We have checked that for such a model we also find spatiotemporally chaotic
behavior, which will be reported elsewhere. 

Spatiotemporally chaotic systems are rare in nonlinear systems in one space dimension,
e.g., the one-dimensional Burgers equation does
not show chaotic behavior.
When described in terms of its intrinsic coordinates our filament
could, naively, considered to be, 
a spatiotemporally chaotic one dimensional system.
However, this is not true because the external shear cannot be captured
only with the intrinsic coordinates. 

A single rigid ellipsoid in a shear flow shows low-dimensional chaotic motion
for appropriate choice of parameters~\cite{radhakrishnan1999numerical,
  lundell2011effect,nilsen2013chaotic}.
Hence, from a fundamental point of view,  it is not a surprise that a deformable
thread in a time-dependent shear
can show spatiotemporally chaotic behaviour.
However, it has never been demonstrated before. 

For a small enough filament, e.g., a single large polymer molecule, thermal
effects that we have ignored, may be important.
We have ignored them for two reasons.
First, in many experimental situations~\cite{liu2018morphological}
the filament is large enough that the thermal fluctuations may not be crucial.
Second, we want to address the fundamental question of emergence of
chaotic behavior due to elastic nonlinearities in the absence of any external
stochastic fluctuations.
We further emphasize that most  strategies of increasing mixing in microfluidics
rely on having a non-zero but small molecular diffusivity -- 
`` For efficient mixing to be achieved, the velocity field must
stir together different portions of the fluid to within a scale that
is small enough for diffusion to take over and homogenize the
concentrations of the advected quantities."~\cite{aref2017frontiers}.
By contrast, we operate at zero molecular diffusivity -- our system is diffusive
even at infinite Peclet number.
%-----------------------------------
\begin{acknowledgments}
  We acknowledge the support of the Swedish Research Council
  Grant No. 638-2013-9243 and 2016-05225.
The figures in this paper are plotted using the free software
matplotlib~\citep{Hunter:2007}.
The simulations were performed on resources provided by
the Swedish National Infrastructure for Computing (SNIC) at PDC center for
high performance computing.
\end{acknowledgments}
\clearpage
\appendix
\section{Model and Method}
\label{sect:method}
\subsection{Model}
%------------------------------
\begin{figure}[!htb]
\begin{center}
  \includegraphics[width=\linewidth]{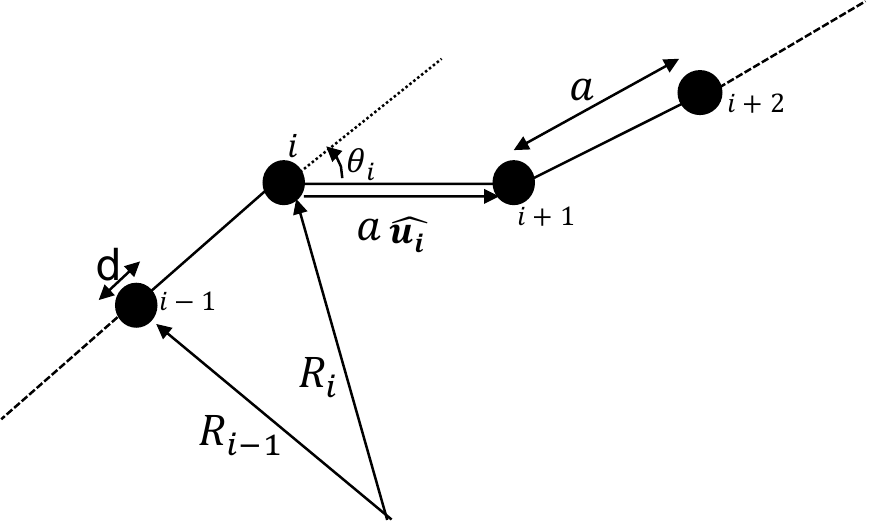}
  \caption{{\bf Schematic} of a freely jointed bead-rod chain. We show $a>d$ for illustration, but
  we use $a=d$ for our simulation.\label{fig:beadrod}
  }
\end{center}
\end{figure}
%------------------------------------
We use the bead-spring model for the numerical simulation of the filament in
a flow \cite{larson1999brownian,guglielmini2012buckling,nazockdast2017fast,
  slowicka2019flexible,
  wada2006non,zuk2021universal}.
The model consists of $N$ spherical beads of diameter $d$,
connected by overdamped springs of equilibrium length $a$, see~\fig{fig:beadrod}. 
The total length of the filament is  $L$.
The position of the center of the $i$-th bead is $\RRi$, where $i = 1,\ldots N$.
The equation of motion for the $i$-th bead is given by~\cite{wada2006non}: 
\begin{equation}
  \delt \Ri^{\alpha} = -\sum_{j=0}^{N-1}M_{\rm ij}^{\alpha\beta}(\Rij)
  \frac{\partial \mH}{\partial \Rj^{\beta}}
  + U^\alpha_0(\RRi) \quad\/.
  \label{eq:dRdtA}
\end{equation}
Where $\Rij = \RRj - \RRi$, $\mH$ is the elastic hamiltonian,
$\partial (\cdot)/\partial(\cdot)$ denotes the partial derivative,
the greek indices run from $1$ to $D$, where $D=3$ is dimensionality of space.
Repeated greek indices are summed.
The velocity of the background flow, $\UU_0$ is given
by
\begin{equation}
  \UU_{0}(x,y) = \gdot y \xhat\/,\quad\text{with}\quad \gdot = S\sin(\omega t)\/,
\label{eq:UU}
\end{equation}
being the time--periodic strain---rate and  $\omega \equiv 2\pi/T$. 
  
The hydrodynamic interaction between the beads is encoded by
the Rotne--Prager mobility tensor $M_{\rm ij}(\RR)$~\cite{rotne1969variational,
  guazzelli2011physical,brady1988stokesian,kim2013microhydrodynamics}:
\begin{widetext}
  \begin{equation}
  M_{\rm ij}^{\alpha\beta}(\RR) =
  \left\{
  \begin{array}{ll}
  \frac{1}{8\pi\eta R}\left[\dab +
  \frac{\Ra\Rb}{R^2} +
  \frac{d^2}{2R^2}\left(\frac{{\dab}}{3} -
  \frac{\Ra\Rb}{R^2}\right)\right] & , {\rm i} \neq {\rm j}  \\
  \frac{1}{3\pi\eta d} \dab & , {\rm i} = {\rm j}
  \end{array}
\right\}.
  \label{eq:mobi}
\end{equation}
\end{widetext}
Here $\eta$ is viscosity of the fluid, and $ R = \left|\RR\right|$.

The Hamiltonian of the system, $\mH$, is $\mH = \mH_{\rm B} + \mH_{\rm S}$
-- we do not consider torsion.
Here $\mH_{\rm B}$ and $\mH_{\rm S}$ are contributions from
bending \cite{montesi2005brownian,bergou2010discrete} and
stretching \cite{wada2006non,wada2007stretching} respectively.
The bending energy of a filament is given by \cite{powers2010dynamics}:
\begin{equation}
  \mH_{\rm B} = \frac{B}{2}\int \kappa^2(s) ds,
\end{equation}
where $B$ is the bending modulus, $\kappa$ is curvature, and $s$ is
the material coordinate.
The discrete form of $\mH_{\rm B}$ is \cite{bergou2008discrete,bergou2010discrete,montesi2005brownian}:
\begin{equation}
  \mH_{\rm B} = aB\sum_{i=0}^{N-1} \kappa_i^2
  = \frac{B}{a}\sum_{i=0}^{N-1} \uuh_i \cdot \uuh_{i-1}
  = \frac{B}{a}\sum_{i=0}^{N-1} \cos \theta_i \/,
    \label{eq:bendingenergy}
\end{equation}
where
\begin{subequations}
  \label{eq:kappa}
\begin{align}
  &\kappa_\ri = \frac{2}{a}\tan\left(\frac{\theta_\ri}{2}\right)
  \approx \frac{\sin\left(\theta_\ri\right)}{a}
  = \frac{\left|\uuh_\ri\times\uuh_{\ri-1}\right|}{a},
  \label{eq:kapa}\\
 &\uuh_\ri = \frac{\RR_{\ri+1} - \RR_\ri}{\lvert{\RR_{\ri+1} - \RR_\ri}\rvert}\quad\/,
\end{align}
\end{subequations}
and $\theta_\ri$ is the angle between two consecutive unit vectors $\uuh_\ri$ and
$\uuh_{\rm {i-1}}$ (see \subfig{fig:snapshot}{A}).
In the second step of \eq{eq:bendingenergy}, we have dropped a constant term.
In the last step of \eq{eq:kapa},
we have used the small-angle approximation \cite{montesi2005brownian}.

The stretching energy is discretized~\cite{wada2006non,wada2007stretching} as: 
\begin{equation}
  \mH_{\rm S} = \frac{H}{2a}\sum_{\ri=0}^{N-1} (\lvert{\RR_{\ri+1} - \RR_\ri}\rvert - a)^2 \/,
\end{equation}
where $H$ is the stretching modulus. We ignore thermal fluctuations.

\subsection{Non--dimensionalization}
We use $L$ and $1/S$ as our characteristic scales
for length and time, respectively.
The evolution equation in non-dimensional form is:.
\begin{multline}
\delttl \Ritl^{\alpha} = -\frac{1}{\mub}\sum_{j=0}^{N-1}\Mtl_{\rm ij}^{\alpha\beta}(\Rijtl)
    \left[\frac{\partial \mHtl_{\rm B}}{\partial \Rjtl^{\beta}} +
    K\left(\frac{L}{d}\right)^3\left(\frac{d}{a}\right) \frac{\partial \mHtl_{\rm S}}{\partial \Rjtl^{\beta}} \right]\\
    + \ytl \sin{\sigma \ttl} \quad\/.
    \label{eq:dRdtAtl}
\end{multline}
Here $\tilde{(\cdot)}$ denotes non-dimensional quantities. 
We get the following dimensionless parameters:
The elasto-viscous number,
\begin{equation}
  \mub \equiv \frac{8\pi\eta S L^4}{B},
\end{equation}
the non-dimensional frequency,
\begin{equation}
  \sigma \equiv \frac{\omega}{S},
\end{equation}
and the ratio of stretching to bending,
\begin{equation}
  K \equiv \frac{H d^2}{B}.
\end{equation}
All the parameter values are shown in table~\ref{table:variable_values}.
\section{Numerical Implementation}
We use the adaptive Runge-Kutta method \cite{press2007numerical}
with cash-karp parameters \cite{press1992adaptive,cash1990variable} to evolve
the system.
We use time-step, $\Delta t$, such that
\begin{equation}
  \tilde{\Delta} = \frac{B\Delta t}{8\pi\eta L^4} = 10^{-11} \text{ -- } 10^{-12}
\end{equation}
We use numerical accuracy of order
$10^{-6}$~\cite{press2007numerical,press1992adaptive,cash1990variable}.
We use CUDA to parallelize the code \footnote{Our code is available here:
  https://github.com/dhrubaditya/ElasticString}.

\begin{figure*}
    \centering
    % \hspace*{-0.3in}
    \includegraphics[width=\textwidth]{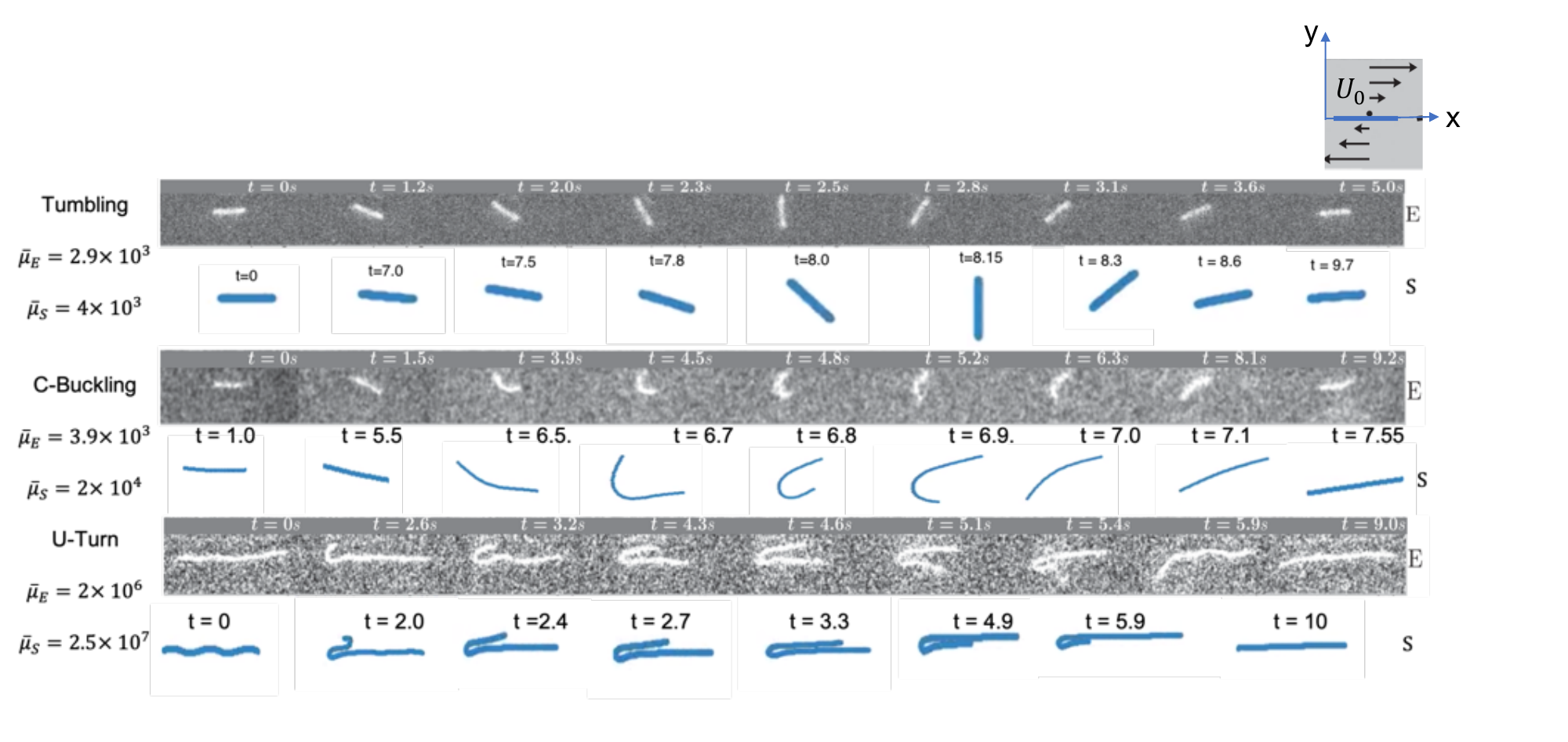}
    \caption{ {\bf Comparison with experimental results:}
      We reproduce the experimental results of Ref.~\cite{liu2018morphological}.
      The filament lies along the x-axis. It is advected by a flow  
      $\UU = (\dot{\gamma}y,0)$, where $\dot{\gamma}$ is the shear rate.
      The flow is constant in time. 
      We observe different dynamical behavior for different $\mub$.
      The grey background are figures from  Ref.~\cite{liu2018morphological}, and the white
      background are results of our simulation.
      Initially, we add small perturbation to the filament
      but ignore thermal fluctuations.
      The  $\mub_{\rm E}$ and $\mub_{\rm S}$ are the values of $\mub$ from
      experiments and our simulations respectively.
      }
    \label{fig:compare}
\end{figure*}

The dimensionless frequency, $\sigma$, must be small enough such that the
Stokesian approximation remains valid.
We use $K=16$ (see table \ref{table:variable_values}) for all the simulations.
Note that \eq{eq:bendingenergy} is exact for an inextensible
filament. In our case, the total length of the filament changes at most
by $2\%$ -- the filament is practically inextensible.
Hence \eq{eq:bendingenergy} remains a reasonable approximation.
Our code reproduces the experimental results by
Liu \etal \cite{liu2018morphological} (see \fig{fig:compare}).
%---------------------------------
\begin{table*}
\centering
\begin{tabular}{ | c | c | }
\hline
\textbf{Parameters} & \textbf{Simulation values}\\
\hline
Number of beads, $N$ & $256$  \\ 
\hline
Equilibrium distance between beads, $a$ & $0.005$  \\ 
\hline
Filament diameter, $d$ & $0.005$\\
\hline
Filament length, $L$ & $1.28$\\
\hline
Bending modulus, $B$ & $2\times10^{-5} \mbox{--} 8\times10^{-3}$\\
\hline
Strain Rate amplitude, $S$  & $2$ \\
\hline
Viscosity, $\eta$ & $10$\\
\hline
Rate of change of strain rate, $\omega$ & $1 \mbox{--} 6$ \\
\hline
Time-step, $\Delta $ & $10^{-4} \mbox{--} 10^{-6}$ \\
\hline
Elasto-viscous number, $\mub = \frac{8\pi\eta S L^4}{B}$ & $1.7 \times 10^5 \mbox{--} 6.8 \times 10^7 $ \\
\hline
Frequency parameter, $\sigma = \frac{\omega}{S}$  & $0.5 \mbox{--} 3$ \\
\hline
Stretching-bending modulus ratio, $K = \frac{H a^2}{B}$ & $16$\\
\hline
\end{tabular}
\caption{Parameters of simulation. Earlier studies have used $N=20-40$~\cite{wada2006non},
  $N=40$~\cite{slowicka2019flexible}, N=400~\cite{chakrabarti2020flexible}}
\label{table:variable_values}
\end{table*}
%----------------------------------------------
%-------------------------------------------------
\section{Detailed description of the dynamical phases}
\label{sect:phase}
Our simulations reveal five different dynamical phases which we call --
straight, buckling, two-period, complex and
complex--transients (see \fig{fig:phase} from main text).

For each case, we describe the dynamics through the morphology of the filament.
This we do in two ways:
\begin{enumerate}[label=(\alph*)]
  \item Extrinsic (real space) -- actual shape of the filament.
  \item Intrinsic (configurational space) -- curvature ($\kappa$) of the filament as a
    function of the material coordinate($s$).
    Our classification of dynamical phase is based on the intrinsic coordinates.
    Conversion between extrinsic to intrinsic coordinates is done using \eq{eq:kappa} -- this conversion is
    unique. Although the inverse is not true. To impose uniqueness, we fix the position of the first bead
    and slope of the bond to the next one.
\end{enumerate}

\subsection{Straight (\textbf{S})}
The filament does not buckle but remains straight all through its
evolution, the curvature remains zero always.

\subsection{Periodic buckling (\textbf{B})}
The filament develops buckling instability. 
The filament settles into periodic behavior after initial transients and
repeats itself stroboscopically (after every cycle) both extrinsically and intrinsically.

\subsection{Two-period (\textbf{2P})}
The filament does not repeat itself after every period but
after every two periods.
Also, the filament does not come back to its position but is rotated
 after two-cycles -- which we call \textit{swimming}.
\subsection{Complex \textbf(C)}
\begin{figure*}
    \includegraphics[width=\textwidth]{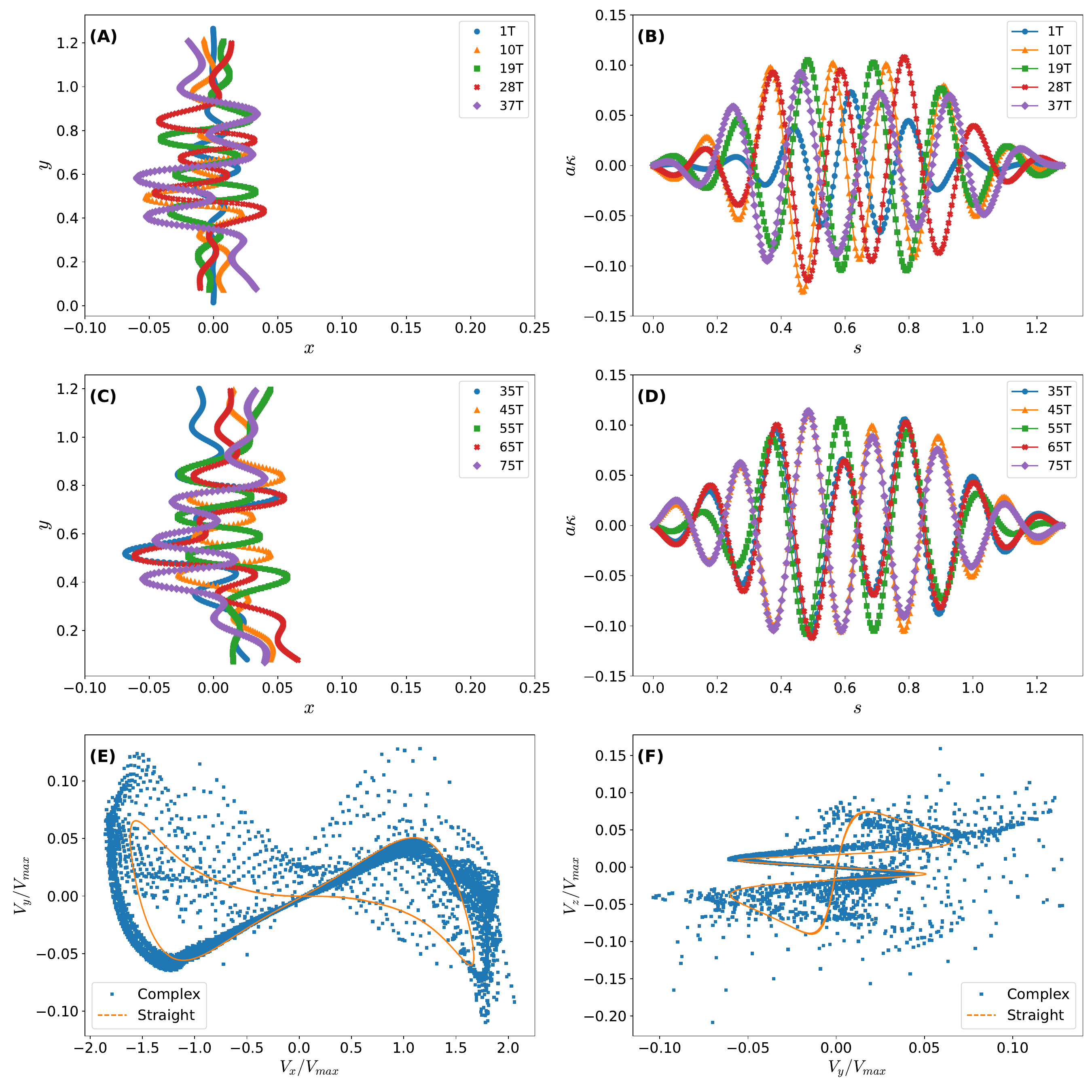}
    % \vspace*{-20mm}
    \caption{{\bf Evolution of the filament in complex phase} ($\mub = 3.35 \times 10^6$, $\sigma = 1.5$) 
    The filament shows complex behavior for cycles at intermediate times, (A) real space;
    (B) configurational space. (C,D) For late cycles also, the filament either does not repeat itself or
      comes close to repeating itself with very high time-period. 
      The filament shows maximum compression at the end of cycle.  
      (E,F) Phase portrait of tracer velocity late cycles $t = 45T$--$75T$.  For comparison
      we also show the  phase potrait for a case in the straight phase (organge).}
    \label{fig:ComplexPhaseEnd}
\end{figure*}

\begin{figure*}
  \includegraphics[width=\textwidth]{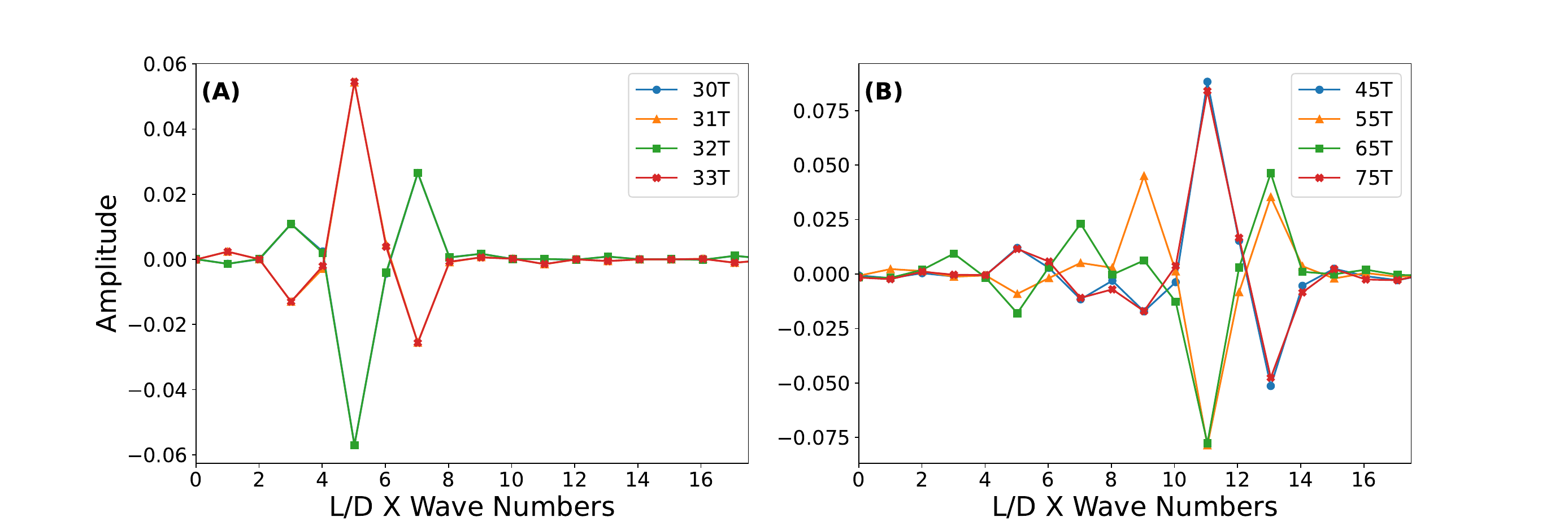}
  \caption{{\bf Discrete sine transform} for two-period phase (A), and complex phase (B).}
\label{fig:DST}
\end{figure*}

\begin{figure*}
    \includegraphics[width=\textwidth]{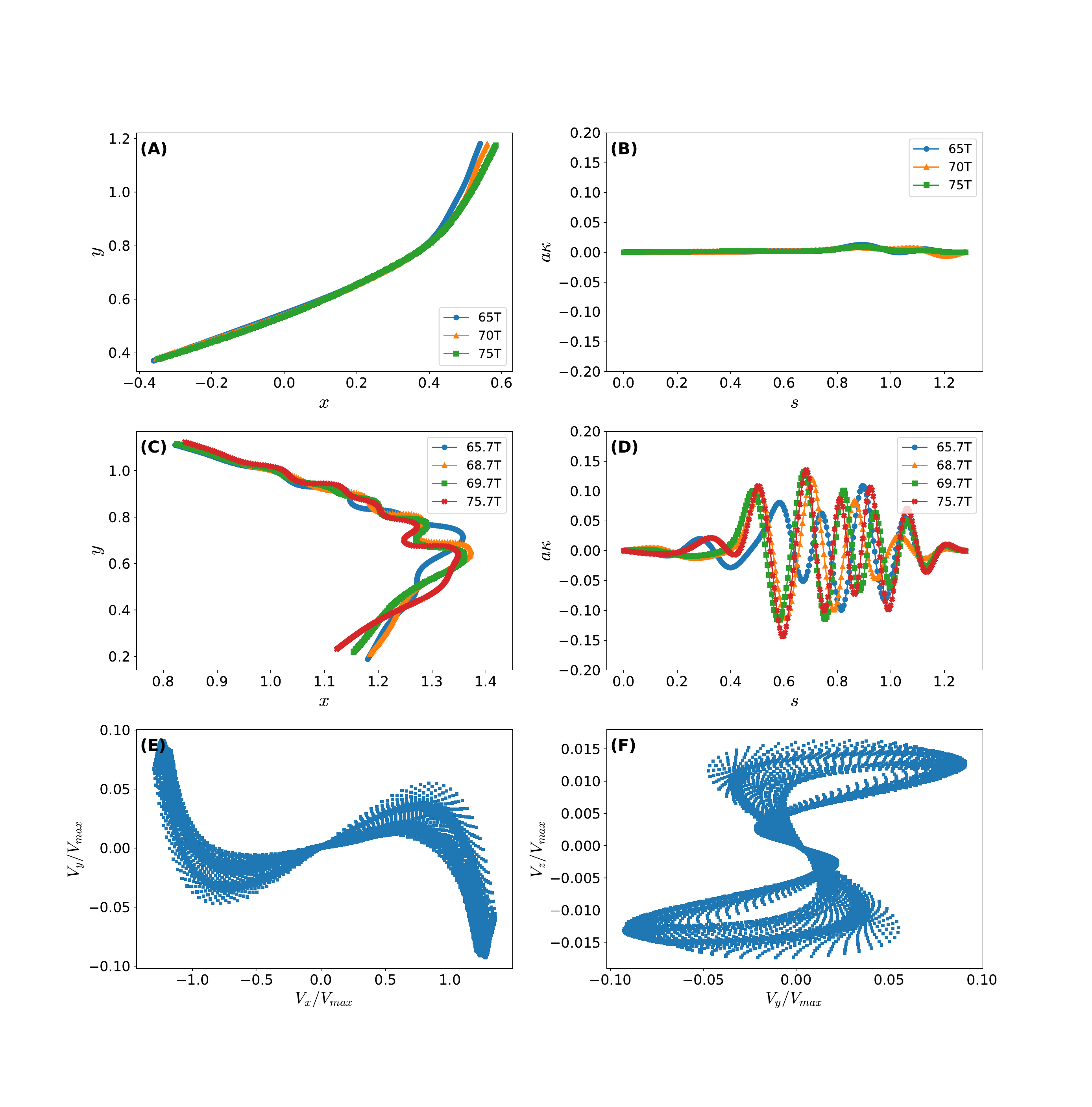}
    \caption{{\bf Evolution of the filament in the complex phase} ($\mub = 6.7 \times 10^6$, $\sigma = 0.75$) The filament shows complex behavior at early cycles respectively in real space and configuration space. 
 However, for late times ($t=65T, 70T, 75T$), the filament almost repeats itself at $nT$, where $n$ is an 
 integer (A,B). Also note that, highest bending energy of the filament is at $(n+p)T$, $p\neq 0$ instead 
 of $nT$ (C,D). The filament is shown stroboscopically for $p=0.7$ respectively in real space and 
 configuration space. We observe that the filament does not repeat itself. 
  (E,F) Phase portrait in $(x,y)$ and $(y,z)$ of Eulerian velocity at one point.
    \label{fig:ComplexPhaseBetween} }
\end{figure*}

The filament rotates in the first half of the cycle (see \fig{fig:snapshot}).
In the second half, it buckles.
In \fig{fig:ComplexPhaseEnd} (A), we plot the filament at the end of $1$st, $10$th, $19$th and $28$th cycle for $\mub = 3.35 \times 10^6$, $\sigma = 1.5$) -- the filament never repeats itself.
In \fig{fig:ComplexPhaseEnd} (B), we plot the curvature ($\kappa$) of the filament as a function of arc-length ($s$) at the same times.
This shows too that the shape of the filament never repeats at the end of each cycle. 
Even at late times $t>60T$, the filament does not repeat itself at the end of a cycle -- see \subfig{fig:ComplexPhaseEnd}{C} where we plot the shape of the filament at $t = 35T, 45T, ..., 75T$. 
The corresponding plot of $\kappa$ versus $s$ is shown in \subfig{fig:ComplexPhaseEnd}{D}.
Here it may seem that the filament comes close to its previous shapes but a careful look tells us that it never completely repeats itself.
Note that, the $\kappa$ -- $s$ plot for $t=35T$ is very close to the one at $t=65T$, although not exactly the same.
The same is true for $t=45T$ and $t=75T$.
This suggests that there maybe a very high period solution to the stroboscopic map.
A sine transform of the $\kappa(s)$ to $\kaph(q)$  shows several peaks,
For $q \gtrapprox 20$ the $\kaph$ is practically zero (\fig{fig:DST})
This demonstrates that our simulations are well resolved to capture the
phenomena we observe. \\
%--------------------------
%----------------------------------
Note that, in some cases of this complex phase, at late times, the filament achieves the most
buckled state (as measured by total elastic energy) not at the end of the cycle but somewhere in between.
One such case is shown in \fig{fig:ComplexPhaseBetween} for $\mub = 6.7 \times 10^6$ and $\sigma = 0.75$.
The $\kappa$--$s$ plot at the end of every cycle comes very close to repeating itself -- 
\subfig{fig:ComplexPhaseBetween}{B}.
The corresponding plots of the filament in real space, is shown in \subfig{fig:ComplexPhaseBetween}{A}, 
are also very close to one another although do not overlap everywhere. 
However, if we loook at intermediate times \eg at $t=65.7T ... 75.7T$, we find that the filament does 
not repeat itself, \subfig{fig:ComplexPhaseBetween}{C,D}. 

To measure the disturbances in the flow due to moving filament, we calculate the time series of 
flow disturbance ($\UU - \UU_0$) at an Eulerian point $\rr = [0,L/2,2d]$.
The numerical method is described in the main body of the paper, see also chapter 8 of 
Ref.~\cite{kim2013microhydrodynamics}.
The Eulerian point is chosen to be just above the $XY$ plane such that the filament does not overlap 
with it.
We show the phase portraits of fluctuating velocity at late times ($t=40T$ to $t=75T$) in 
\fig{fig:ComplexPhaseEnd}, \subfig{fig:ComplexPhaseBetween}{E,F} respectively.
Note that, the velocity values are larger compared to the 
straight phase (\subfig{fig:ComplexPhaseEnd}{E,F}).
%------------------------------------------%
%----------------------------------------%
\subsection{Complex transients:}
\label{subsect:CT}
\begin{figure*}
  \includegraphics[width=0.95\linewidth]{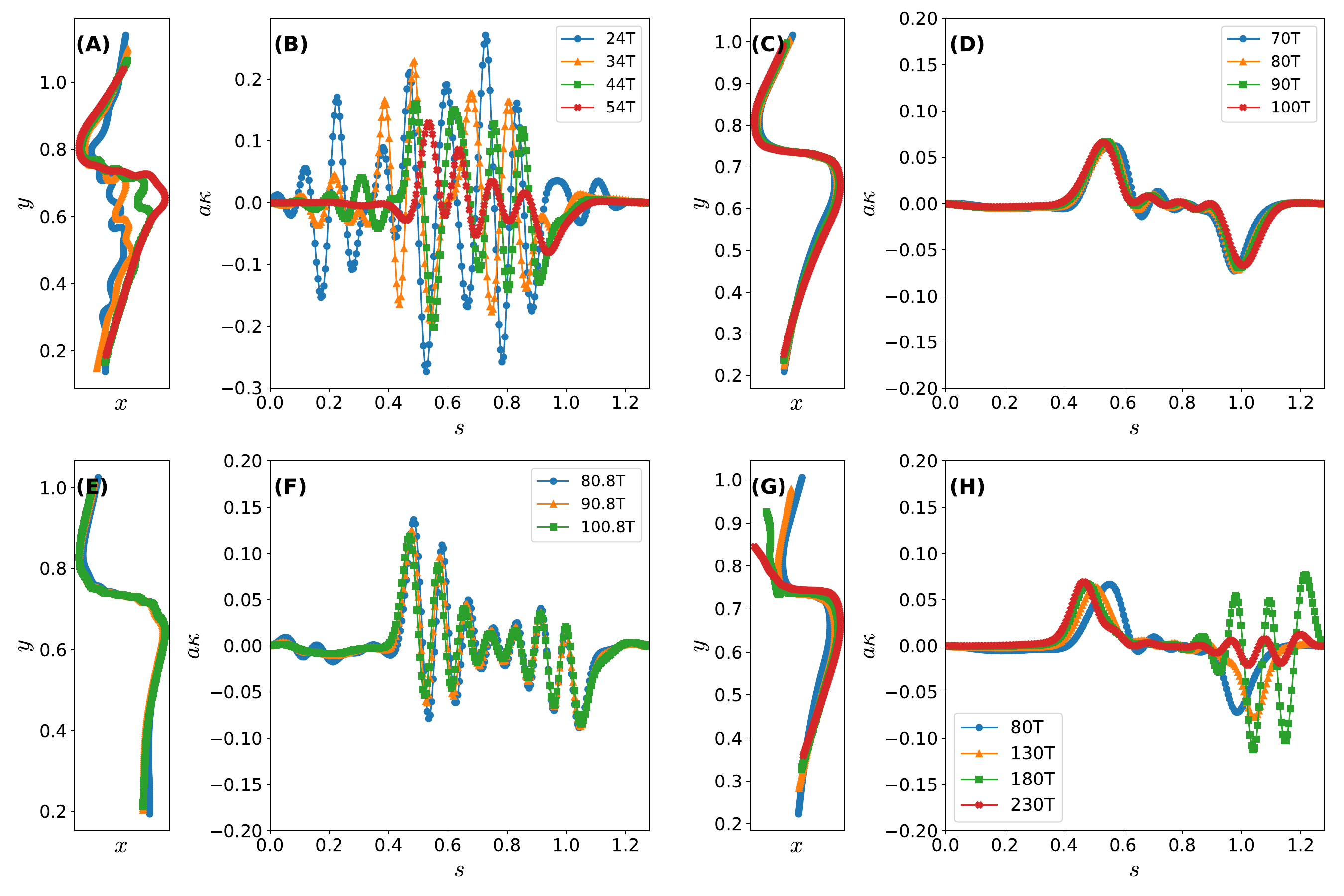}
  \includegraphics[width=\linewidth]{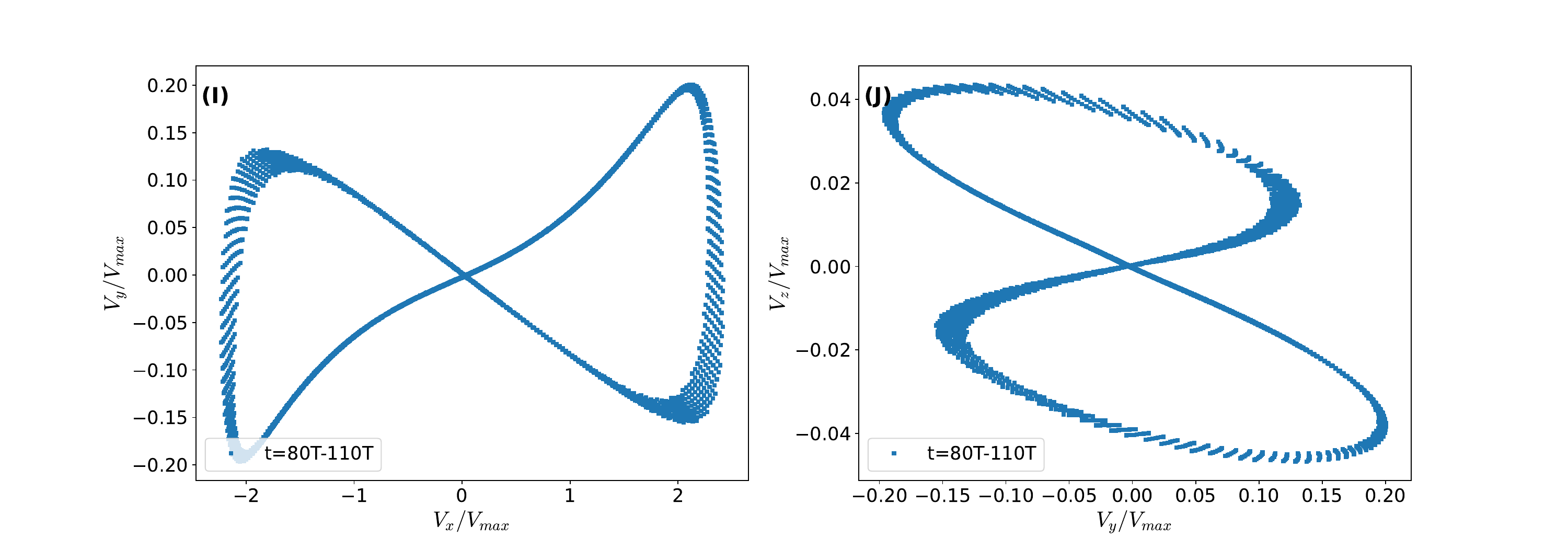}
  \caption{{\bf Evolution in the complex--transient phase}: 
($\mub = 3.35 \times 10^7$, $\sigma = 1.5$). The shape of the filament in (A)(C)(E) and the 
corresponding curvature is shown in (B)(D)(F). For early cycles (t $\le$ 60T), the filament 
shows complex behavior and does not repeat itself (A,B), similar to the complex 
phase~(see \fig{fig:ComplexPhaseEnd}).
For late cycles (t=70T \dots 100T), although the filament does not repeat itself, 
it comes very close after every cycle (C,D,E,F). 
The filament shape, from the end of one cycle to next, changes very slowly,
\eg (E) shows the shape at $t=80.8T,90.8T,100.8T$.
In (G), we show the shapes at $t=80T,130T \dots$.
Over such a long time scale, the shape does change.
(I)(J) Phase portrait of velocity at an Eulerian point for the late cycles.}
\label{fig:ComplexTransient}
\end{figure*}

The filament shows high mode of buckling.
We compare the filament extrinsically and intrinsically at the end of $24$th, $34$th, $44$th 
cycle for $\mub = 3.35 \times 10^7, \sigma = 1.5$ respectively 
in~\subfig{fig:ComplexTransient}{A,B}.
The filament shows complex behavior and does not repeat itself for early periods ($t<60T$), similar to the complex phase.
However, the complex behavior is transient and the filament comes very close to itself for 
late periods -- at the end of a cycle ($t=80T,90T,100T$, see \subfig{fig:ComplexTransient}{C})
and intermediate times between cycle where the filament shows maximum 
buckling ($t=80.8T,90.8T,100.8T$, see \subfig{fig:ComplexTransient}{E}).
The corresponding plots of $\kappa$--$s$ are shown in \fig{fig:ComplexTransient} (D)(F) -- 
this also shows the almost--periodic behavior of filament at late times.
In \subfig{fig:ComplexTransient}{G}, we show the shapes at $t=80T,130T, 180T$ and $230T$.
Over such a long time scale, the shape does change.
The corresponding $s$--$\kappa$ plots are shown in \subfig{fig:ComplexTransient}{H} 
In \subfig{fig:ComplexTransient}{I} and (J) we show the phase portrait 
 of Eulerian velocity at $\rr = [0,L/2,2d]$ for late times ($t>80T$).
%-----------------------------------------------------%
%-----------------------------------------%
\section{Stroboscopic map}
\label{sect:map}
We take a dynamical system approach to analyze the complex dynamics we observe.
Such techniques have been used widely to study highly turbulent
flows~\cite{kerswell2005recent,suri2020capturing,page2020exact}
and has also been applied to other fields of fluid
dynamics~\cite{franco2018bubble,gaillard2021life}.
Let us define an operator $\cF$ such that:
\begin{subequations}
  \begin{align}
    &\Kkappa(T) = \cF \Kkappa(0)\\
    &\cF^\rp = \cF \cF ... \text{p times} ... \cF,
  \end{align}
\end{subequations}
where $\Kkappa = [\kappa_1,\kappa_2, \dots \kappa_\ri \dots \kappa_N]$, where $\kappa_\ri$ is 
the curvature at point $\ri$.
For a given $\Kkappa$, the operator, $\cF$, returns the values of $\Kkappa$ after evolving 
the system for exactly one time-period.
We look for fixed points and periodic orbits of this map~\cite{cvitanovic2005chaos}
by solving 
$\Kkappa = \cF^\rp \Kkappa$. 
The task is now to calculate the solutions of set of the non-linear equations:
\begin{equation}
  \label{eq:cN}
  \cN_\rp\Kkappa \equiv \left(\cF^\rp -1\right)\Kkappa = 0
\end{equation}
For example, $\Kkappa = \bm{0}$, $\rp = 1$, is a solution for straight phase.
The periodic buckling and two--period have non--zero curvature ($\Kkappa$) solution 
respectively for $\rp = 1$ and $\rp = 2$.
We use the Newton-Krylov method~\cite{knoll2004jacobian} based on Generalized Minimal
Residual Method~(GMRES)~\cite{saad1986gmres} in jacobian--free way to find the solutions.
It is described next.
\subsection{Newton--Krylov}
\label{subsect:NK}
%------------------------------
\begin{figure}
    \includegraphics[width=0.9\columnwidth]{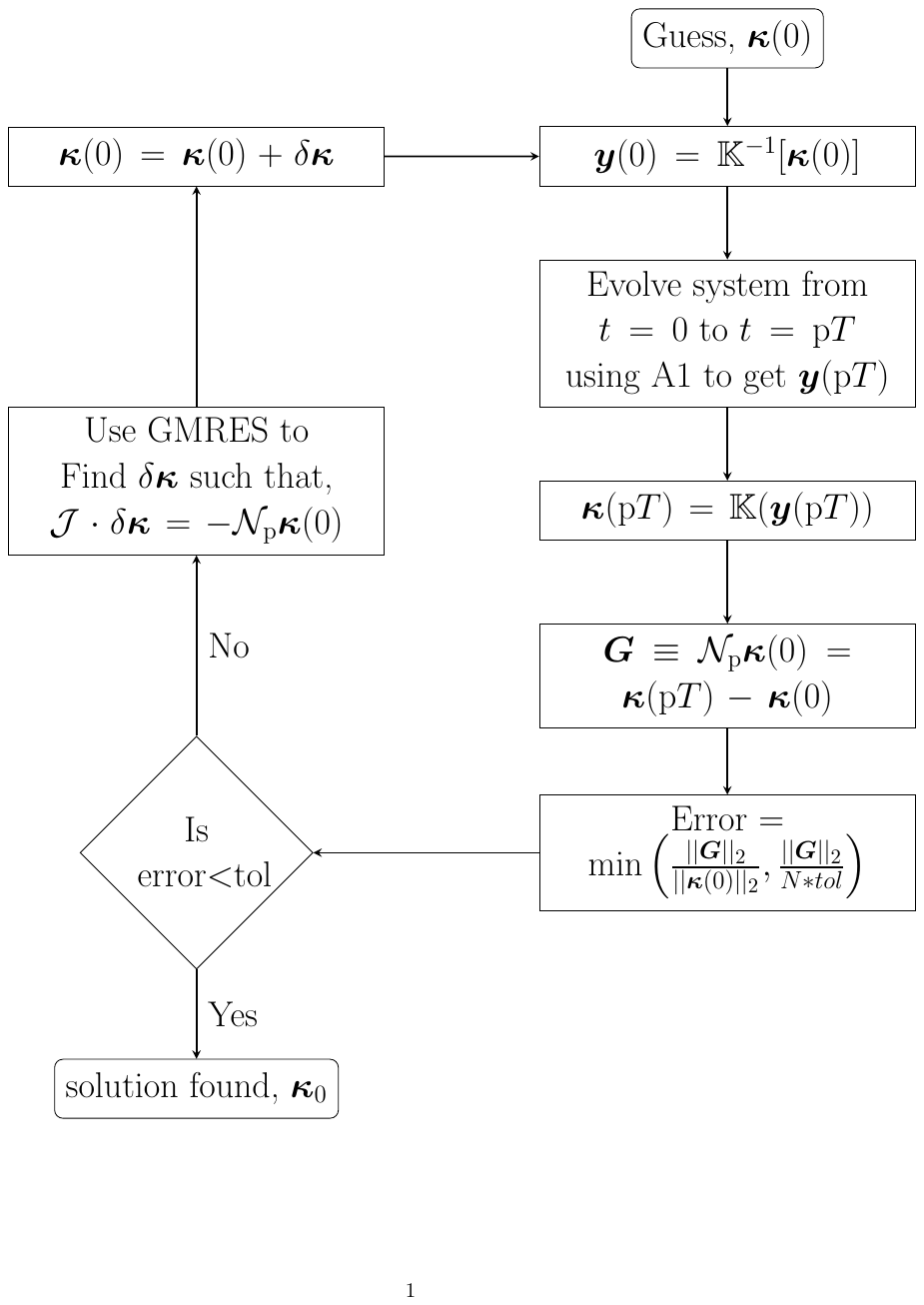}
    \caption{Flowchart for Newton--Krylov iteration. $\Kbb$ is co-ordinate transformation
      from real to curvature space using equation~\eq{eq:kappa}.
      Similarly, $\Kbb^{-1}$ is the inverse coordinate transformation from curvature to
      real space. We use the notation: $\cJ$ is jacobian of the operator
      $\cN$, described in \eq{eq:cN}.
      We use $\rm{tol}=0.01$ }
    \label{fig:flowchart}
\end{figure}
%-----------------------------
The flow-chart of the algorithm is shown in \fig{fig:flowchart}.
We start with a guess for the curvature, $\Kkappa(0)$.
Then we calculate the positions of the beads given by
$\yy\equiv[y_1, ... y_{2j-1},y_{2j}, ... y_{2N}] \equiv [R_1^1,..., R_j^1,R_j^2,...R_N^2]$.
We call this transformation $\Kkappa$ to $\yy$, $\Kbb^{-1}$, such that:
\begin{equation}
  \yy(0) = \Kbb^{-1}\Kkappa(0)
\end{equation}
Then we evolve in time from $t=0$ to $t=\rp T$ by solving \eq{eq:dRdtA} to obtain 
$\yy(\rp T)$.
Then we apply the inverse transformation, $\Kbb$, to obtain
\begin{equation}
  \Kkappa(\rp T) = \Kbb (\yy (\rp T))
\end{equation}
Then we check how accurately \eq{eq:cN} is satisfied, \ie we define
\begin{equation}
  \text{error} = \frac{||\cN_\rp\Kkappa||_2}{||\Kkappa(0)||_2}.
\end{equation}
The case for straight solution is dealt specially because in this case $\Kkappa = 0$. 
Here we use:
\begin{equation}
  \text{error} = \frac{||\cN_\rp\Kkappa||_2}{N*\text{tol}}
\end{equation}
If the error is less than a preset tolerance, then we accept the guess ($\Kkappa(0)$) as 
a solution, otherwise, we generate a new guess by
\begin{subequations}
  \begin{equation}
  \Kkappa(0) = \Kkappa(0)+\delta\Kkappa,
  \end{equation}
  \text{such that}
  \begin{equation}
  \label{eq:AX=B}
  \cJ\cdot\delta\Kkappa = -\cN_\rp\Kkappa(0).
  \end{equation}
\end{subequations}
Here $\cJ$ is jacobian matrix of the operator $\cN_\rp$ computed at $\Kkappa(0)$.
We do not calculate $\cJ$, instead we calculate:
% We calculate:
\begin{equation}
  \label{eq:jacobianvectorproduct}
  \cJ\cdot\delta\Kkappa = \frac{\cN_\rp(\Kkappa(0)+\varepsilon\delta\Kkappa) 
                - \cN_\rp(\Kkappa(0)-\varepsilon\delta\Kkappa)}{2\varepsilon}.
\end{equation}
Here $\varepsilon$ is a step size \cite{knoll2004jacobian}.
We use the GMRES \cite{saad1986gmres,guennebaud2010eigen} method in matrix-free way
using \eq{eq:jacobianvectorproduct} to find solutions of \eq{eq:AX=B}.
The operator $\cN_\rp$ is implemented as described in \Fig{fig:flowchart}.
The value of $\varepsilon$ should be small enough such that \eq{eq:jacobianvectorproduct} is
well approximated and large enough such that the floating point round--off errors are not
too large~\cite{knoll2004jacobian}.
We compute $\varepsilon$ in the following way:
\begin{equation}
  \varepsilon = \varepsilon_{\rm {rel}}
  \left(1+\frac{||\Kkappa(0)||_2}{||\delta\Kkappa(0)||_2}\right),
\end{equation}
where $||\cdot||$ is the $2$nd norm, and $\varepsilon_{\rm {rel}} = 10^{-4}$.

Note that, the conversion from curvature space to real space ($\Kbb^{-1}$) is not unique.
However, if we fix the position of the first bead and slope of the bond to the next one, 
it is unique.
One advantage of using this method is that, it accounts for all the continuous
symmetries (translation in $x$,$y$ direction) present in the
system~\cite{cvitanovic2005chaos}.

Also note that the $\Kkappa$ is the same for two filaments which has the same shape but
are rotated with respect to each other.
But the evolution of two such filaments are not the same, in principle, because
the ambient flow depends on space.
In some cases, we find the solutions such that the filament comes to the same shape
as before but rotated.
We call these solutions ``swimming solutions''.
\end{document}